\documentclass[%
superscriptaddress,
preprint,
tightenlines,
showpacs,
preprintnumbers,
showkeys,
nofootinbib,
amsmath,
amssymb,
aps,
prd,
]{revtex4-1}
\usepackage{CJK}
\usepackage{verbatim}
\usepackage{multirow}
\usepackage{dcolumn}
\usepackage{bm}
\usepackage{graphicx}
\usepackage[caption=false]{subfig}
\usepackage[x11names,svgnames]{xcolor}

\usepackage{hyperref}
\usepackage{bookmark}

\newcommand{\Nxyz}{N_{\text x} N_{\text y} N_{\text z}}

\newcommand{\csw}{{c_{\text{sw}}}}

\DeclareMathOperator{\tr}{tr}
\DeclareMathOperator{\re}{Re}
\DeclareMathOperator{\im}{Im}
\DeclareMathOperator{\dd}{d\!}

\begin{document}

\begin{CJK*}{UTF8}{}

\preprint{UTHEP-660}  \preprint{UTCCS-P-74}  \preprint{KANAZAWA-13-07}

\title{Finite size scaling study of $N_{\text{f}}=4$ finite density QCD on the lattice}

%% Before the final submission to APS, all \CJKfamily macro calls
%% should be commented out.
\author{Xiao-Yong Jin
  (\CJKfamily{bsmi}金\CJKkern{}曉\CJKkern{}勇)}
\affiliation{
  RIKEN Advanced Institute for Computational Science,
  Kobe, Hyogo 650-0047, Japan
}

\author{Yoshinobu Kuramashi
  (\CJKfamily{min}藏\CJKkern{}増~嘉\CJKkern{}伸)}
\affiliation{
  Faculty of Pure and Applied Sciences,
  University of Tsukuba, Tsukuba, Ibaraki 305-8571, Japan
}
\affiliation{
  Center for Computational Sciences,
  University of Tsukuba, Tsukuba, Ibaraki 305-8577, Japan
}
\affiliation{
  RIKEN Advanced Institute for Computational Science,
  Kobe, Hyogo 650-0047, Japan
}

\author{Yoshifumi Nakamura
  (\CJKfamily{min}中\CJKkern{}村~宜\CJKkern{}文)}
\affiliation{
  RIKEN Advanced Institute for Computational Science,
  Kobe, Hyogo 650-0047, Japan
}

\author{Shinji Takeda
  (\CJKfamily{min}武\CJKkern{}田~真\CJKkern{}滋)}
\affiliation{
  Institute of Physics,
  Kanazawa University, Kanazawa 920-1192, Japan
}
\affiliation{
  RIKEN Advanced Institute for Computational Science,
  Kobe, Hyogo 650-0047, Japan
}

\author{Akira Ukawa
  (\CJKfamily{min}宇\CJKkern{}川~彰)}
\affiliation{
  Center for Computational Sciences,
  University of Tsukuba, Tsukuba, Ibaraki 305-8577, Japan
}
\affiliation{
  RIKEN Advanced Institute for Computational Science,
  Kobe, Hyogo 650-0047, Japan
}

\date{\today}

\begin{abstract}
  We explore the phase space spanned by the temperature and the
  chemical potential for 4-flavor lattice QCD using the Wilson-clover
  quark action.  In order to determine the order of the phase
  transition, we apply finite size scaling analyses to gluonic and
  quark observables including plaquette, Polyakov loop and quark
  number density, and examine their susceptibility, skewness, kurtosis
  and Challa-Landau-Binder cumulant.  Simulations were carried out on
  lattices of a temporal size fixed at $N_{\text{t}}=4$ and spatial
  sizes chosen from $6^3$ up to $10^3$.  Configurations were generated
  using the phase reweighting approach, while the value of the phase
  of the quark determinant were carefully monitored.  The
  $\mu$-parameter reweighting technique is employed to precisely
  locate the point of the phase transition.  Among various
  approximation schemes for calculating the ratio of quark
  determinants needed for $\mu$-reweighting, we found the Taylor
  expansion of the logarithm of the quark determinant to be the most
  reliable.  Our finite-size analyses show that the transition is
  first order at $(\beta, \kappa, \mu/T)=(1.58, 0.1385, 0.584\pm
  0.008)$ where $(m_\pi/m_\rho, T/m_\rho)=(0.822, 0.154)$.  It weakens
  considerably at $(\beta, \kappa, \mu/T)=(1.60, 0.1371, 0.821\pm
  0.008)$ where $(m_\pi/m_\rho, T/m_\rho)=(0.839, 0.150)$, and a
  crossover rather than a first order phase transition cannot be ruled
  out.  %(JIN: Estimated uncertainties for physical scales?)
\end{abstract}

% \pacs{
%   11.10.Wx \jinnote{not found in 2010 version?},
%   % 11.15.Ha, % Lattice gauge theory
%   % 12.38.-t, % Quantum chromodynamics
%   % 12.38.Aw, % General properties of QCD (dynamics, confinement, etc.)
%   12.38.Gc % Lattice QCD calculations
%   % 12.38.Mh, % Quark-gluon plasma,
%   % 25.75.Nq % phase transitions in Quark-gluon plasma,
% }

% \keywords{
%   \jinnote{keywords} Lattice QCD, Finite temperature and density
% }

\maketitle
\end{CJK*}

%% \listoffixmes

\tableofcontents

\section{Introduction}

The 4-flavor QCD is a good testing ground for finite temperature and
chemical potential analyses before studying the physically more
relevant case of the 3-flavor theory.  In fact, since the 4-flavor
theory can be described with the staggered fermion formalism without
rooting, new ideas to explore QCD with finite density have first been
tried out in this
theory~\cite{Fodor:2001au,DElia:2002gd,Kratochvila:2005mk}.

\begin{figure}
  \subfloat[]{\includegraphics[scale=0.32]{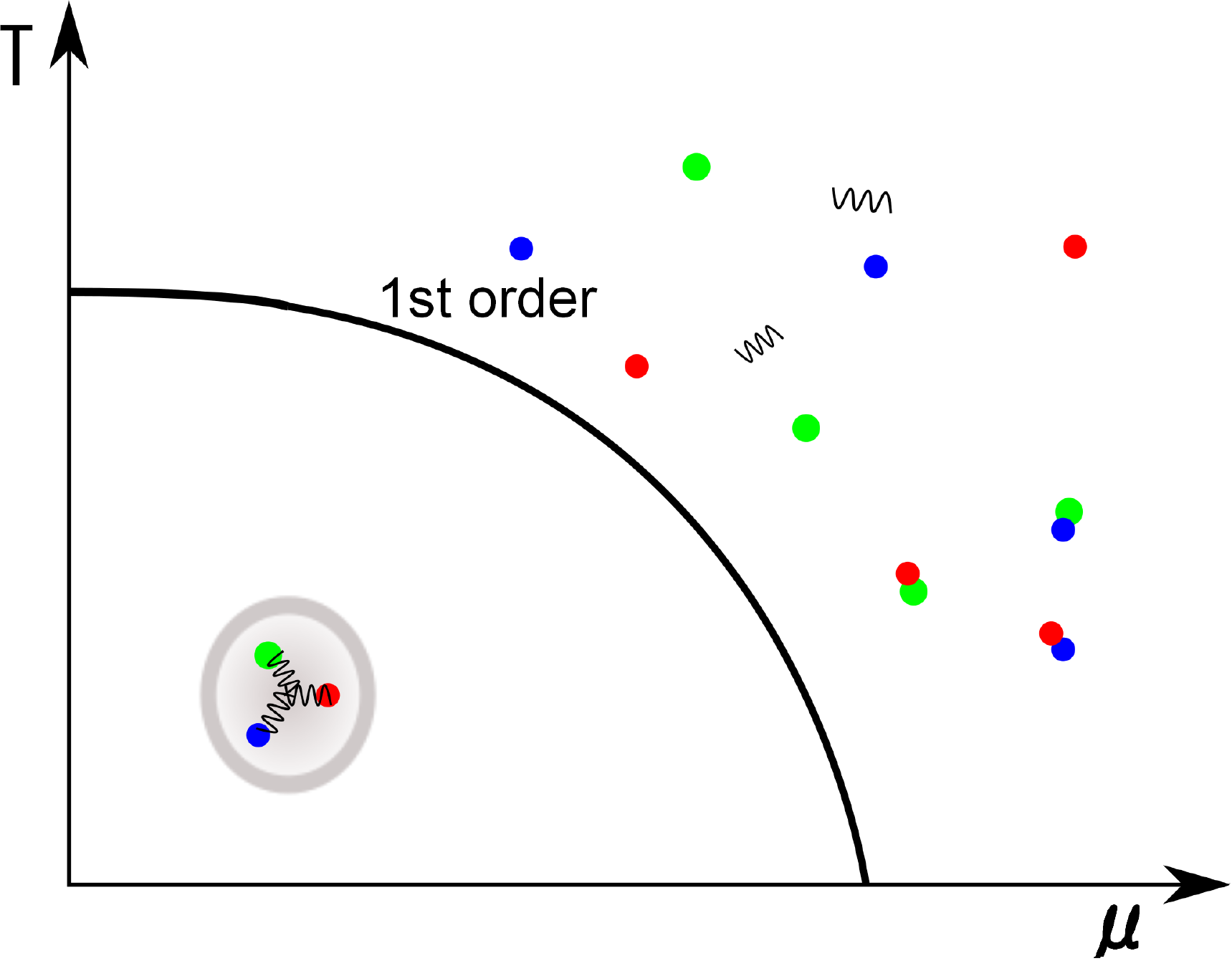}} \qquad
  \subfloat[]{\includegraphics[scale=0.32]{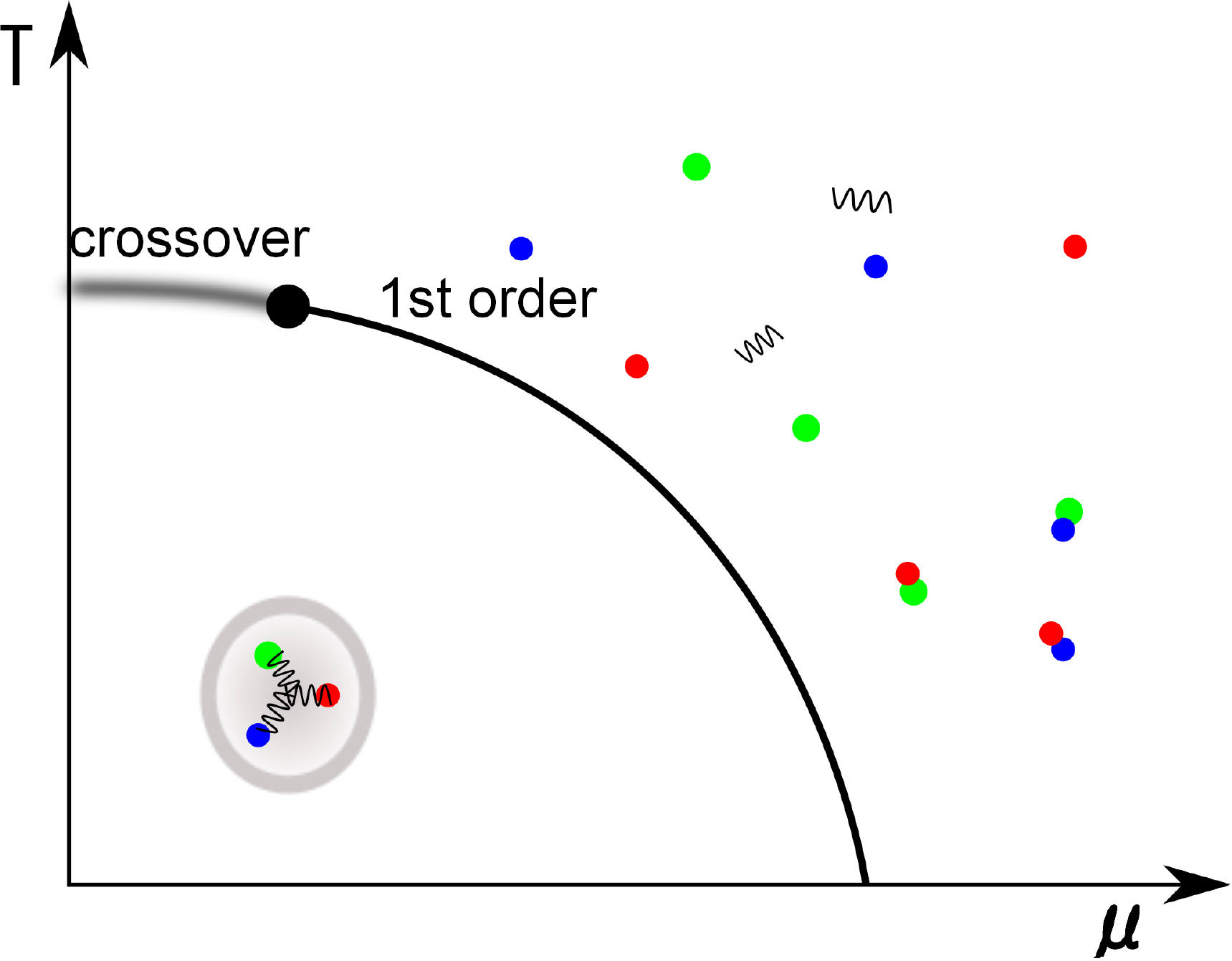}}
  \caption{\label{fig:phasediagram} Phase diagram on the $(\mu, T)$
    plane expected for 4-flavor QCD for (a) massless quarks, and (b)
    heavy quarks. }
\end{figure}

More fundamentally, the phase diagram of the 4-flavor theory is
expected to have a structure well suited for exploratory studies at
finite density.  With massless quarks, as shown in
Fig.~\ref{fig:phasediagram}(a), a continuous line of first order phase
transitions connects the temperature and chemical potential axes.
When the quark mass, $m_q$, is increased, the first order phase
transition at zero density turns into a crossover beyond some value of
$m_q$, while the transition at zero temperature and finite density
remains first order as shown in Fig.~\ref{fig:phasediagram}(b).  Consequently the first order line up to some
value of the chemical potential also turns into a crossover.  Hence a
critical end point is expected at a finite chemical potential, which
is reminiscent of the situation for the 3-flavor theory with the
physical spectrum of up, down and strange quarks.  It is empirically
known~\cite{Fukugita:1990vu,Iwasaki:1995ij} in the zero density case
that the first order phase transition persists up to a relatively
large quark mass in the 4-flavor theory.  Therefore one should be able
to probe the region of the transition line with a reasonable
computational cost, and learn much about the physical characteristics
of the transition before tackling a more difficult 3-flavor theory.

A powerful method for resolving the nature of phase transition is the
finite size scaling analysis.  While this method has been extensively
exploited in lattice QCD studies at finite temperatures, the situation
appears quite different at non-zero baryon density.  This is partly
due to the fact that, in the phase-reweighting procedure for numerical
%%% YN
%simulations at non-zero density, the average phase of the quark
%determinant is expected to decrease exponentially as the lattice
simulations at non-zero density, the averaged phase-reweighting factor
 is expected to decrease exponentially as the lattice
%%% YN
volume increases, leading to a loss of control of statistical averages
of observables.  In addition the calculation of the quark determinant
necessary for evaluating the phase is computationally very expensive.

We note, however, that the former problem does not necessarily
%%% YN
%preclude finite-size scaling analyses as long as the phase average
preclude finite-size scaling analyses as long as the reweighting factor
%%% YN
stays reasonably away from zero over the range of lattice volumes
needed for the analysis.  This is a dynamical question, and as we have
%%% YN
%shown in Ref.~\cite{Takeda:2011vd} the average phase becomes larger
shown in Ref.~\cite{Takeda:2011vd} the averaged phase-reweighting factor becomes larger
%%% YN
for larger temporal lattice sizes.  Concerning the latter, the
reduction of the quark determinant~\cite{Danzer:2008xs,Nagata:2010xi}
and the recent development of computing technology including high
speed GPGPU have significantly extended the range of lattice sizes for
which the determinant is calculable in practice.  In this article we
therefore make a serious attempt at finite size scaling analyses for
non-zero density QCD.

The Kentucky group~\cite{Li:2010qf} studied the phase structure of the
4-flavor theory using the canonical approach employing the
Wilson-clover quark action.  They observed an S-shaped structure in
the chemical potential versus quark number plot, which they took to be
an indication of a first order phase transition.  The study was only
on a single lattice volume of $6^3\times4$ and with relatively low
statistics, however, so this may not be taken as a conclusive
statement.  From the point of view of universality, it is important to
check the phase structure by using different approaches.  Accordingly,
we also employed the Wilson-clover quark action, but adopted the grand
canonical approach, and performed a finite size scaling study to learn
how we can quantitatively resolve the order of the transition.

The rest of the paper is organized as follows.  We briefly discuss the
phase reweighting method and parameter reweighting for $\mu$ in
Sec.~\ref{sec:phasereweighting} and \ref{sec:mureweighting},
respectively.  Simulation parameters are summarized in
Sec.~\ref{sec:simulationparameters}.  After defining the observables
we measure in Sec.~\ref{sec:defquantities}, we present our finite size
scaling analysis using susceptibility, skewness, kurtosis and the
Challa-Landau-Binder cumulant for a variety of gluonic and quark
observables in Sec.~\ref{sec:finitedensitymoment}.  By combining with
results of zero density simulation, we describe a sketch of the global
phase diagram in Sec.~\ref{sec:phasediagram}.  In the last section, we
present our concluding remarks.  In
Appendix~\ref{sec:doublepeakdistribution} we summarize an analysis of
volume scaling of higher moments by using a double Gaussian
distribution model, and in
Appendix~\ref{sec:mureweighting_for_quark_number} some details of
$\mu$-reweighting for observables which explicitly depend on $\mu$ are
given.

Throughout this paper we consider a 4-dimensional Euclidean lattice of
a size specified by $N_{\text x}\times N_{\text y}\times N_{\text
  z}\times N_{\text t}$.  The boundary condition is periodic in the
spatial directions, while in the temporal direction, it is periodic
(anti-periodic) for gluon (quark) fields.  Some preliminary results
given in this paper were already reported at the Lattice 2012
Conference~\cite{Takeda:2012ar}.

\section{\label{sec:phasereweighting}Phase reweighting}

Physics of QCD for finite quark chemical potential $\mu$ can be
studied by the grand canonical partition function. Assuming that the
$N_{\text f}$ quark flavors are degenerate, {\it i.e.}, all quarks
have the same mass and chemical potential, the partition function is
given by
\begin{align}
  \mathcal{\mathcal{Z}}_{\text{QCD}}(\mu)
  &= \int[\dd U]e^{-S_{\text G}}\det D(\mu)^{N_{\text f}} \,,\\
  &= \int[\dd U]\exp (-S_{\text{QCD}}) \,, \\
  S_{\text{QCD}} &= S_{\text G} - N_{\text f} \ln \det D(\mu) .
\end{align}
We adopt the Wilson-clover quark action with the Wilson-Dirac matrix,
\begin{align}
  D(\mu) &= \delta_{x,y} -\kappa \sum_{\nu=1}^4 [
  e^{\mu a\delta_{\nu,4}} (1-\gamma_\nu)U(x,\nu) \delta_{x+\hat\nu,y}
  +e^{-\mu a\delta_{\nu,4}}
  (1+\gamma_\nu)U(y,\nu)^\dag\delta_{x-\hat\nu,y} ]
  \nonumber\\
  &+ \kappa \csw \delta_{x,y}
  F_{\nu\rho}(x)\sigma_{\nu\rho}
\end{align}
%% (ukawa: is the sign in front of \mu correct? ST:corrected)
with $\mu$ the chemical potential, $a$ the lattice spacing and
$F_{\nu\rho}(x)$ the standard clover term.  We employ the Iwasaki
gauge action~\cite{Iwasaki:1983}
%% (ukawa: the definition of iwasaki action may be dropped as suggested by nakamura san. ST I keep it.)
\begin{equation}
  \begin{split}
    S_{\text G} = \beta \sum_x \Big\{& c_0 \sum_{\nu<\rho} \big(
    1-\frac{1}{3} \re W_{\nu\rho}^{1\times1}(x) \big) \\
    &+ c_1 \sum_{\nu,\rho} \big( 1-\frac{1}{3} \re
    W_{\nu\rho}^{1\times2}(x) \big) \Big\},
  \end{split}
\end{equation}
with $c_1=-0.331$, $c_0=1-8c_1=3.648$, and the gauge invariant loops
are given by
\begin{align}
  W_{\nu\rho}^{1\times1}(x) = \tr \big[&
  U(x,\nu)U(x+\hat{\nu},\rho) \nonumber \\
  & U(x+\hat{\rho},\nu)^\dag U(x,\rho)^\dag \big],
  \label{eqn:plaquetteloop} \\
  W_{\nu\rho}^{1\times2}(x) = \tr \big[&
  U(x,\nu)U(x+\hat{\nu},\rho)U(x+\hat{\nu}+\hat{\rho},\rho) \nonumber \\
  & U(x+2\hat{\rho},\nu)^\dag U(x+\hat{\rho},\rho)^\dag U(x,\rho)^\dag
  \big] \,.
\end{align}

Since the quark determinant with $\mu\neq0$ is complex,
% \begin{equation}
%\det D(\mu)^\dag
%=
%\det D(-\mu),
%\end{equation}
one cannot apply the standard Monte Carlo simulation.  Defining the
phase of the quark determinant with
\begin{equation}
  \label{eq:63}
  \det D(\mu)
  \equiv
  |\det D(\mu)| e^{i\theta(\mu)},
\end{equation}
one can rewrite the expectation value of an observable $\mathcal{O}$ as
\begin{equation}
  \langle
  \mathcal{O}
  \rangle
  =
  \frac
  {\langle \mathcal{O}e^{iN_{\text f}\theta}\rangle_{||}}
  {\langle e^{iN_{\text f}\theta}\rangle_{||}},
  \label{eqn:reweighting}
\end{equation}
where the phase-included and the phase-quenched ensemble averages are
given by
\begin{align}
  \langle \mathcal{O} \rangle &= \frac{ \int [\dd U] e^{-S_{\text G}}(\det
    D)^{N_{\text f}}\mathcal{O}[U] }{ \int [\dd U] e^{-S_{\text G}}(\det
    D)^{N_{\text f}}
  }, \\
  \langle \mathcal{O} \rangle_{||} &= \frac{ \int [\dd U] e^{-S_{\text
        G}}|\det D|^{N_{\text f}}\mathcal{O}[U] }{ \int [\dd U] e^{-S_{\text
        G}}|\det D|^{N_{\text f}}
  }\,. \label{eqn:phasequenchedaverage}
\end{align}
This defines the phase-reweighting method, which allows evaluation of
%%% YN
%observables as long as the average phase ${\langle e^{iN_{\text
observables as long as the averaged phase-reweighting factor ${\langle e^{iN_{\text
%%% YN
      f}\theta}\rangle_{||}}$ stays non-zero.  In general this
factor vanishes exponentially with the space-time lattice volume,
leading to the sign problem.  In practice, however, the numerical
%%% YN
%magnitude of the average phase is dynamically determined.  Hence
magnitude of the averaged phase-reweighting factor is dynamically determined.  Hence
%%% YN
viability of the phase-reweighting method can only be determined by
actual simulations.  Furthermore, we have shown in
%%%% YN
%%Ref.~\cite{Takeda:2011vd} that the average phase increases for larger
Ref.~\cite{Takeda:2011vd} that the averaged phase-reweighting factor increases for larger
%%% YN
temporal lattice sizes, with other parameters fixed in the heavy quark
mass region.  Therefore we expect that the phase-reweighting method
provides information on the phase structure over practically useful
parameter region.

Another practical issue of the phase-reweighting method is how to
compute the phase factor which requires a computationally expensive
calculation of the determinant.  In order to avoid introduction of
systematic errors, we perform an exact calculation of the quark
determinant by adopting the reduction technique of
Ref.~\cite{Danzer:2008xs}.  After reduction in the temporal direction,
the quark determinant can be expressed as
\begin{align}
%  \det D(\mu/T)
  \det D(\mu)
  &= A_0 W(\mu/T) \nonumber \\
  &= A_0 \det \left[1 - H_0 - e^{\mu/T}H_+ - e^{-\mu/T}H_-\right] \,,
  \label{eqn:WHHH}
\end{align}
where the definition of $A_0$, $H_\pm$ and $H_0$ are given in
Ref.~\cite{Takeda:2011vd}.  After numerically building $H_{\pm}$ and
$H_0$ which are dense matrices of order $12\Nxyz$, the determinant in
Eq.~\eqref{eqn:WHHH} can be computed by using the LU decomposition.
We also perform a reduction in the spinor space.  In total the number
of floating point operations for calculating the determinant is
reduced by about a factor of two compared to the non-reduced case.  In
our simulations we exploit GPGPU to carry out the determinant
calculation in the reduced form.
%% ukawa:can you quote the original flop and the reduced flop to show roughly how much computer time reduction is achieved, N_t^2??? ST I quote a factor.)

\section{\label{sec:mureweighting}$\mu$-reweighting}
%In this section we remind the $\mu$-reweighting
%which is very useful in identifying and studying the phase transition point.
%In course of implementing $\mu$-reweighting, one has to rely on an approximation
%to avoid a huge computational cost.
%We will examine some expansion schemes by using actual simulation data in Sec.~%\ref{sec:simul-results-from-mu}.

%\subsection{Preliminary}

In finite size scaling analyses we often need to calculate the
position of extrema of moments of observables.  Since they are usually
not located at the points of simulation, reweighting methods as
originally proposed in Ref.~\cite{Ferrenberg:1989ui} are very useful.
In our case, we want to evaluate physical quantities at a chemical
potential $\mu^\prime$ from phase quenched configurations generated at
a value $\mu\neq\mu^\prime$.  For this purpose, we can use the
identity,
\begin{align}
\langle
{\cal O}(\mu^\prime)
\rangle_{\mu^\prime}
=
\frac{
\left\langle
{\cal O}(\mu^\prime)
\frac{\det D(\mu^\prime)^{N_{\text f}}}{\det D(\mu)^{N_{\text f}}}
e^{iN_{\text f}\theta(\mu)}
\right\rangle_{||\mu}
}
{
\left\langle
\frac{\det D(\mu^\prime)^{N_{\text f}}}{\det D(\mu)^{N_{\text f}}}
e^{iN_{\text f}\theta(\mu)}
\right\rangle_{||\mu}
},
\label{eqn:mureweighting}
\end{align}
where the phase-quenched average at $\mu$ in the right hand side is
defined in Eq.~\eqref{eqn:phasequenchedaverage}.

%A key quantity for this reweighting is the ratio of determinant and
A practical question here is how to evaluate the ratio of quark
determinants.  Due to its huge computational cost, we have to avoid a
direct computation of the full determinant at each reweighted value of
the chemical potential.  Instead we exploit an approximation to the
determinant, and introduce three expansion schemes: winding expansion,
Taylor expansion of the determinant, and Taylor expansion of the
logarithm of the determinant.

As shown in Eq.~\eqref{eqn:WHHH}, the $\mu$ dependence of the
determinant is factorized, and $A_0$ does not appear in the ratio of
the determinants.
\begin{equation}
  \frac{\det D(\mu^\prime)^{N_{\text f}}}{\det D(\mu)^{N_{\text f}}}
  =
  \frac{W(\mu^\prime/T)^{N_{\text f}}}{W(\mu/T)^{N_{\text f}}}.
\label{eqn:ratio}
\end{equation}
In the following we consider only $W(\mu/T)$.

%\subsection{Expansion schemes of fermion determinant}

The winding expansion~\cite{Danzer:2008xs} is an expansion of $\log W(\mu/T)$
in terms of fugacity $\exp \mu/T$;
\begin{equation}
  W(\mu/T)
  =
  \exp
  \left[
    -V\sum_{q\in\mathbb{Z}}
    v^{(q)}e^{q\mu/T}
  \right],
  \label{eqn:Wdefwinding}
\end{equation}
where the lattice spatial volume $V$ is factored out in the argument.
%In the paper \cite{Takeda:2011vd}, the author denoted $V^{(q)}=Vv^{(q)}$.
In an actual implementation, one has to truncate the expansion at some order
$q=q_{\text{trunc}}$.  The approximated form of the ratio is given by
\begin{align}
  \frac{\det D(\mu^\prime)^{N_{\text f}}}{\det D(\mu)^{N_{\text f}}}
  &\longrightarrow \exp \left[ -N_{\text f}V\sum_{q=1}^{q_{\text{trunc}}} 2 \re
    [v^{(q)}] \left\{ \cosh(q\mu^\prime/T)-\cosh(q\mu/T) \right\}
  \right.
  \nonumber\\
  & \phantom{\exp[} \left.  -iN_{\text f}V\sum_{q=1}^{q_{\text{trunc}}} 2 \im
    [v^{(q)}] \left\{ \sinh(q\mu^\prime/T) - \sinh(q\mu/T) \right\}
  \right].
\end{align}
The second line is considered as an additional phase difference
between two fermion determinants.  The $v^{(q)}$'s are constructed
from $H_0$ and $H_\pm$ in Eq.~\eqref{eqn:WHHH}.  In practice we choose
$q_{\text{trunc}}=10$.

% \subsubsection{Taylor expansion of determinant}

In order to define Taylor expansions, we introduce two types of derivatives,
$Q_n$ defined by
\begin{equation}
  Q_n
  =
  \frac{1}{W^{N_{\text f}}}
  \frac{\partial^n W^{N_{\text f}}}{\partial (\mu/T)^n},
\end{equation}
and $W_n$ by
\begin{equation}
  \frac{\partial^n \ln W^{N_{\text f}}}
  {\partial (\mu/T)^n}
  =
  N_{\text f}W_n.
  \label{eqn:defWn}
\end{equation}
These two derivatives can be related to each other as moments and
their cumulants.  Up to $n=1,2,3,4$ the relations take the form,
\begin{subequations}
  \label{eqn:Q}
  \begin{align}
    Q_1 &= N_{\text f}W_1,
    \label{eqn:Q1}
    \\
    Q_2 &= N_{\text f}W_2 + (N_{\text f}W_1)^2,
    \label{eqn:Q2}
    \\
    Q_3 &= N_{\text f}W_3 + 3 (N_{\text f}W_2) (N_{\text f}W_1) +
    (N_{\text f}W_1)^3,
    \label{eqn:Q3}
    \\
    Q_4 &= N_{\text f}W_4 + 4 (N_{\text f}W_3) (N_{\text f}W_1) + 3
    (N_{\text f}W_2)^2 + 6 (N_{\text f}W_2) (N_{\text f}W_1)^2 +
    (N_{\text f}W_1)^4,
    \label{eqn:Q4}
  \end{align}
\end{subequations}
and the explicit form of $W_n$'s are given by
\begin{subequations}
  \label{eqn:W}
  \begin{align}
    W_1 &= \tr [B],
    \label{eqn:W1}
    \\
    W_2 &= - \tr [B^2] + \tr [C],
    \label{eqn:W2}
    \\
    W_3 &= 2 \tr [B^3] - 3 \tr [BC] + \tr [B],
    \label{eqn:W3}
    \\
    W_4 &= -6 \tr [B^4] + 12 \tr [B^2C] - 4 \tr [B^2] - 3 \tr [C^2] +
    \tr [C],
    \label{eqn:W4}
    \\
    B &= K^{-1}\frac{\partial K}{\partial (\mu/T)},
    \\
    C &= K^{-1}\frac{\partial^2 K}{\partial (\mu/T)^2},
    \\
    K(\mu/T) &= 1-H_0-H_+e^{\mu/T}-H_-e^{-\mu/T}.
  \end{align}
\end{subequations}
By using $H_0$ and $H_\pm$, one can calculate $W_n$ and $Q_n$.

The Taylor expansion of the ratio of determinants is given by
\begin{align}
  \frac{\det D(\mu^\prime)^{N_{\text f}}}{\det D(\mu)^{N_{\text f}}}
%  &= \sum_{n=0}^\infty \frac{ (\mu^\prime/T-\mu/T)^n }{n!}
%  \frac{1}{\det D(\mu)^{N_{\text f}}} \frac{\partial^n \det
%    D(\mu)^{N_{\text f}}}{\partial (\mu/T)^n}
%  \nonumber\\
%  &= \sum_{n=0}^\infty \frac{ (\mu^\prime/T-\mu/T)^n }{n!}
%  \frac{1}{W(\mu/T)^{N_{\text f}}} \frac{\partial^n W(\mu/T)^{N_{\text
%        f}}}{\partial (\mu/T)^n}
%  \nonumber\\
  &= 1+ \sum_{n=1}^\infty \frac{ (\mu^\prime/T-\mu/T)^n }{n!}  Q_n.
\end{align}
Note that the $Q_n$ are evaluated at $\mu$.  In our actual
implementation, we truncate the sum at $n=4$.

% \subsubsection{Taylor expansion of logarithm of determinant}
The Taylor expansion of the logarithm of the
determinant ratio is given by
\begin{align}
  \frac{\det D(\mu^\prime)^{N_{\text f}}}{\det D(\mu)^{N_{\text f}}}
%  &= \exp \left[ \ln \det D(\mu^\prime)^{N_{\text f}} - \ln \det
%    D(\mu)^{N_{\text f}} \right]
%  \nonumber\\
%  &= \exp \left[ \sum_{n=1}^\infty \frac{ (\mu^\prime/T-\mu/T)^n
%    }{n!}  \frac{\partial^n \ln \det D(\mu)^{N_{\text f}}}{\partial
%      (\mu/T)^n} \right]
%  \nonumber\\
%  &= \exp \left[ \sum_{n=1}^\infty \frac{ (\mu^\prime/T-\mu/T)^n
%    }{n!}  \frac{\partial^n \ln W(\mu/T)^{N_{\text f}}}{\partial
%      (\mu/T)^n} \right]
%  \nonumber\\
  &= \exp \left[ \sum_{n=1}^\infty \frac{ (\mu^\prime/T-\mu/T)^n
    }{n!}  N_{\text f}W_n \right],
\end{align}
%The pure imaginary part of the $W_n$ is considered as an additional phase.
The difference of the phase at $\mu$ and $\mu^\prime$ is given by
\begin{align}
  \theta(\mu^\prime) %&=
  % \im [\ln \det D(\mu^\prime)] =
%  \im [\ln W(\mu^\prime/T)]
%  \nonumber\\
%  &= \sum_{n=0}^\infty \frac{(\mu^\prime/T-\mu/T)^n}{n!}  \im
%  \frac{\partial^n \ln W(\mu/T)}{\partial (\mu/T)^n}
%  \nonumber\\
  &= \theta(\mu) + \sum_{n=1}^\infty
  \frac{(\mu^\prime/T-\mu/T)^n}{n!}  \im W_n.
\end{align}
Practically we truncate the sum at $n=4$.

Since the determinant is a product of eigenvalues of the Wilson-Dirac
matrix whose number grows proportional to lattice volume, we expect
the Taylor expansion of the logarithm of the determinant ratio to be
better behaved toward larger volume than the expansion of the
determinant ratio itself.  We verify this explicitly in
Sec.~\ref{sec:simul-results-from-mu} in our numerical simulations.

For observables which explicitly depends on $\mu$, {\it e.g.}, quark
number density and related quantities, the observables themselves also
have to be evaluated at reweighted values of $\mu$.  In this study
Taylor expansion is used for such observables and the details are
given in Appendix \ref{sec:mureweighting_for_quark_number}.

\section{\label{sec:simulationparameters}Simulation parameters}
In our simulations, we used the clover coefficient $\csw$ calculated
from the formula
\begin{equation}
  \csw
  =
  1+0.113(6/\beta)+0.0209(6/\beta)^2+0.0047(6/\beta)^3.
\end{equation}
It was non-perturbatively determined for the case of $N_{\text
  f}=3$~\cite{Aoki:2005et}.  Nevertheless, we chose it for the present
exploratory study of the $N_{\text f}=4$ case.  This choice also
facilitates a comparison with the work of the Kentucky
group~\cite{Li:2010qf} who adopted the same $\csw$.

%To carry out finite size scaling analyses, the
%spatial volume is varied from $6^3$ to $10^3$ as given in detail later,
%while the temporal lattice size is fixed at $N_{\text t}=4$ in our
%simulations.

We performed non-zero density simulations as well as zero density
ones.  For the non-zero density case, we chose two sets of parameters:
$(\beta,\kappa)=(1.58,0.1385)$ and $(1.60,0.1371)$.  The second set is
exactly the same as that of the Kentucky group~\cite{Li:2010qf}.  The
spatial volume and the chemical potential are summarized in
Table~\ref{tab:parameters_1} for $(\beta,\kappa)=(1.58,0.1385)$ and in
Table~\ref{tab:parameters_2} for $(1.60,0.1371)$.  We chose five
spatial volumes, $6^3$, $6^2\times8$, $6\times8^2$, $8^3$ and $10^3$
for finite size scaling analyses, while fixing the temporal size to
$N_{\text t}=4$.  Our control parameter for the quark number is the
chemical potential and our ensembles cover a range of
$a\mu=0.02-0.35$.  The onset of the charged pion condensate is
expected at $a\mu_c(T=0)=am_\pi/2$.  According to the hadron spectrum
results summarized in Table~\ref{tab:hadronspectrum}, we estimate
$a\mu_c\approx0.65$, and hence we do not need to worry about it in our
parameter region.

For zero density simulations, we chose two sets of parameters:
$(\beta,\kappa)=(1.60,0.1380)$ and $(1.618,0.1371)$, and the spatial
volume was varied from $6^3$ to $12^3$ while $N_{\text t}=4$ was fixed
for both sets.  Simulation parameters are summarized in
Table~\ref{tab:zerodensityparameters}.

We used the BQCD code~\cite{Nakamura:2010qh} which implements the HMC
algorithm and several techniques.  We used the multi-time-scale
technique~\cite{Sexton:1992nu} with a ratio of step sizes of
$d\tau_g:d\tau_d:d\tau_f=1:2:4$ where $d\tau_g$, $d\tau_d$ and
$d\tau_f$ are step sizes for gauge force, logarithm of determinant for
clover term and pseudo-fermion force, respectively.  The Omelyan
integrator~\cite{Takaishi:2005tz} was adopted in our simulation.  In
order to generate a probability distribution containing the
phase-quenched quark determinant, we used the finite iso-spin chemical
potential $\mu_{\text u}=-\mu_{\text d}$.  Two independent
pseudo-fermions were employed to incorporate $N_{\text f}=4$ dynamical
quarks.  We set the trajectory length to unity and fixed the step size
$d\tau_f=1/20$, with which the HMC acceptance rate stayed around
$90$\% for all parameter sets.  For each parameter set,
$20,000-1,200,000$ trajectories were accumulated.  The acceptance rate
and the number of trajectories were compiled in
Tables~\ref{tab:parameters_1} and \ref{tab:parameters_2}.  The
ingredients of the determinant in Eq.~\eqref{eqn:W} were measured at
every 10 trajectories.  We employed jackknife analyses with varying
bin sizes, and chose the maximum estimated statistical error to be
quoted in this paper.

\begin{table}
  \caption{\label{tab:parameters_1}Simulation parameters and
    statistics at $\beta=1.58$ and $\kappa=0.1385$.}
  \begin{ruledtabular}
    \begin{tabular}{c|l|c|r}
      $\Nxyz$ & $a\mu$ & accep. & traj.    \\
      \hline
      $6^3$   & $0.02$ & $0.94$ & $20000$  \\
              & $0.04$ & $0.94$ & $20000$  \\
              & $0.06$ & $0.94$ & $20000$  \\
              & $0.08$ & $0.94$ & $20000$  \\
              & $0.10$ & $0.94$ & $50000$  \\
              & $0.12$ & $0.94$ & $50000$  \\
              & $0.13$ & $0.94$ & $50000$  \\
              & $0.14$ & $0.94$ & $50000$  \\
              & $0.15$ & $0.94$ & $50000$  \\
              & $0.16$ & $0.94$ & $50000$  \\
              & $0.18$ & $0.95$ & $50000$  \\
              & $0.20$ & $0.95$ & $20000$  \\
              & $0.22$ & $0.95$ & $20000$  \\
              & $0.24$ & $0.95$ & $20000$  \\
              & $0.26$ & $0.95$ & $20000$  \\
              & $0.28$ & $0.95$ & $20000$  \\
              & $0.30$ & $0.95$ & $20000$  \\
      \hline
      $668$   & $0.13$ & $0.93$ & $50000$  \\
              & $0.14$ & $0.93$ & $50000$  \\
              & $0.15$ & $0.93$ & $50000$  \\
              & $0.16$ & $0.93$ & $50000$  \\
      \hline
      $688$   & $0.13$ & $0.92$ & $50000$  \\
              & $0.14$ & $0.92$ & $130000$ \\
              & $0.15$ & $0.92$ & $130000$ \\
              & $0.16$ & $0.92$ & $50000$  \\
      \hline
      %       & $0.10$ & $0.91$ & $50000$  \\
      %       & $0.12$ & $0.91$ & $50000$  \\
      $8^3$   & $0.13$ & $0.91$ & $275000$ \\
              & $0.14$ & $0.91$ & $275000$ \\
              & $0.15$ & $0.91$ & $275000$ \\
              & $0.16$ & $0.91$ & $50000$  \\
      %       & $0.18$ & $0.92$ & $50000$  \\
      \hline
      $10^3$  & $0.13$ & $0.87$ & $50000$  \\
              & $0.14$ & $0.87$ & $347800$ \\
              & $0.15$ & $0.87$ & $342800$ \\
              & $0.16$ & $0.87$ & $113900$ \\
      %       & $0.17$ & $0.88$ & $50000$  \\
    \end{tabular}
  \end{ruledtabular}
\end{table}

\begin{table}
  \caption{\label{tab:parameters_2}Simulation parameters and
    statistics at $\beta=1.60$ and $\kappa=0.1371$.}
  \begin{ruledtabular}
    \begin{tabular}{c|l|c|r}
      $\Nxyz$ & $a\mu$  & accep. & traj.     \\
      \hline
      $6^3$   & $0.10$  & $0.95$ & $20000$   \\
              & $0.15$  & $0.95$ & $80000$   \\
              & $0.16$  & $0.95$ & $80000$   \\
              & $0.17$  & $0.95$ & $80000$   \\
              & $0.18$  & $0.95$ & $80000$   \\
              & $0.19$  & $0.95$ & $80000$   \\
              & $0.20$  & $0.95$ & $160000$  \\
              & $0.205$ & $0.95$ & $160000$  \\
              & $0.21$  & $0.95$ & $160000$  \\
              & $0.215$ & $0.95$ & $80000$   \\
              & $0.22$  & $0.95$ & $80000$   \\
              & $0.23$  & $0.95$ & $80000$   \\
              & $0.24$  & $0.96$ & $40000$   \\
              & $0.25$  & $0.95$ & $40000$   \\
              & $0.30$  & $0.96$ & $20000$   \\
              & $0.35$  & $0.96$ & $20000$   \\
      \hline
      $668$   & $0.205$ & $0.94$ & $320000$  \\
      \hline
      $688$   & $0.205$ & $0.93$ & $900000$  \\
      \hline
      $8^3$   & $0.10$  & $0.93$ & $20000$   \\
              & $0.15$  & $0.92$ & $100000$  \\
              & $0.16$  & $0.92$ & $100000$  \\
              & $0.17$  & $0.92$ & $100000$  \\
              & $0.18$  & $0.92$ & $100000$  \\
              & $0.19$  & $0.92$ & $500000$  \\
              & $0.20$  & $0.92$ & $900000$  \\
              & $0.205$ & $0.92$ & $1200000$ \\
              & $0.21$  & $0.92$ & $900000$  \\
              & $0.215$ & $0.92$ & $500000$  \\
              & $0.22$  & $0.93$ & $500000$  \\
              & $0.23$  & $0.93$ & $100000$  \\
              & $0.24$  & $0.93$ & $100000$  \\
              & $0.25$  & $0.93$ & $100000$  \\
              & $0.30$  & $0.93$ & $20000$   \\
              & $0.35$  & $0.93$ & $20000$   \\
    \end{tabular}
  \end{ruledtabular}
\end{table}

\begin{table}
  \caption{\label{tab:hadronspectrum}Hadron spectrum for $N_{\text f}=4$ QCD.}
  \begin{ruledtabular}
    \begin{tabular}{c|c|c||c|c|c}
      $\beta$ & $\Nxyz \times N_{\text t}$        & $\kappa$ & $am_\pi$     & $am_\rho$    & $am_N$       \\
      \hline
      $1.580$ & $12^3\times24$ & $0.1380$ & $1.3666(16)$ & $1.6550(26)$ & $2.6529(39)$ \\
      $1.580$ & $12^3\times24$ & $0.1385$ & $1.3317(16)$ & $1.6197(23)$ & $2.5745(46)$ \\
      $1.580$ & $12^3\times24$ & $0.1390$ & $1.2896(16)$ & $1.5830(23)$ & $2.5108(29)$ \\
      \hline
      $1.600$ & $12^3\times24$ & $0.1371$ & $1.3958(15)$ & $1.6639(25)$ & $2.6473(36)$ \\
      $1.600$ & $12^3\times24$ & $0.1380$ & $1.3275(10)$ & $1.6097(19)$ & $2.5790(42)$ \\
      $1.600$ & $12^3\times24$ & $0.1390$ & $1.2392(15)$ & $1.5340(26)$ & $2.4170(20)$ \\
      \hline
      $1.618$ & $12^3\times24$ & $0.1371$ & $1.3497(19)$ & $1.6166(27)$ & $2.5521(21)$ \\
      $1.618$ & $12^3\times24$ & $0.1380$ & $1.2686(17)$ & $1.5465(31)$ & $2.4810(61)$ \\
      $1.618$ & $12^3\times24$ & $0.1390$ & $1.1511(16)$ & $1.4240(24)$ & $2.2651(42)$ \\
    \end{tabular}
  \end{ruledtabular}
\end{table}

\begin{table}
  \caption{\label{tab:zerodensityparameters}Simulation parameters
    and statistics at $a\mu=0$}
  \begin{ruledtabular}
    \begin{tabular}{c|c|c|c|r}
      $\beta$ & $\kappa$ & $\Nxyz$ & accep. & traj.   \\
      \hline
      $1.600$ & $0.1380$ & $6^3$   & $0.95$ & $40000$ \\
      &&$668$   & $0.94$ & $40000$ \\
      &&$688$   & $0.93$ & $40000$ \\
      &&$8^3$   & $0.92$ & $40000$ \\
      &&$10^3$  & $0.89$ & $40000$ \\
      &&$12^3$  & $0.86$ & $20000$ \\
      \hline
      $1.618$ & $0.1371$ & $6^3$   & $0.96$ & $20000$ \\
      &&$668$   & $0.95$ & $40000$ \\
      &&$688$   & $0.94$ & $40000$ \\
      &&$8^3$   & $0.93$ & $40000$ \\
      &&$10^3$  & $0.91$ & $40000$ \\
      &&$12^3$  & $0.88$ & $20000$ \\
    \end{tabular}
  \end{ruledtabular}
\end{table}

\section{\label{sec:defquantities}Definition of physical
  quantities}

\subsection{\label{sec:moments} Moments and cumulants}
Let $X$ be the space-time average of a local observable.  In general
non-central moments $\mu_n, n=1,2,3,\cdots$ and cumulants $\kappa_n$
of $X$ can be defined by the QCD partition function in the presence of
source term $\mathcal{Z}_{\text{QCD}}(\alpha)=\langle \exp (\alpha X
)\rangle$ according to
\begin{equation}
\mu_n=\left. \frac{1}{\mathcal{Z}_{\text{QCD}}(\alpha)}
\frac{\partial^n \mathcal{Z}_{\text{QCD}}(\alpha) }{\partial \alpha^n}\right|_{\alpha=0},
\end{equation}
and
\begin{equation}
\kappa_n=\left. \frac{ \partial^n \log\mathcal{Z}_{\text{QCD}}(\alpha) }{\partial \alpha^n}\right|_{\alpha=0}.
\end{equation}
%%(ukawa: takeda san, the following sentence should answer your point 1. ST OK I see.)
If the parameter $\alpha$ is contained in the action, one can take the
derivative without introducing the source term.  This applies to the
gluon action density for which $\alpha$ can be taken as the inverse
gauge coupling $\beta$ and the quark number density for which
$\alpha=\mu/T$, apart from some coefficient proportional to volume.

The quantities of the most interest for our finite size scaling
analyses are susceptibility $\chi_X$, skewness $S_X$, and kurtosis
$K_X$ defined respectively by
\begin{eqnarray}
\chi_X&=&V\kappa_2, \\
S_X&=&\frac{\kappa_3}{\kappa_2^{3/2}},\\
K_X&=&\frac{\kappa_4}{\kappa_2^2}.
\end{eqnarray}
We also analyze the CLB (Challa-Landau-Binder)
cumulant~\cite{Fukugita:1989yw,Challa:1986sk}) defined in terms of
non-central moments according to
\begin{equation}
U_X=1-\frac{\mu_4}{3\mu_2^2}.
\end{equation}

Divergence of the susceptibility peak height with volume is a
well-known indicator of the nature of the transition.  Both the peak
of the susceptibility and the zero of the skewness $S_X=0$ can be
interpreted as the location of the transition point.
%Note that the kurtosis is slightly different from the 4-th
%order Binder cumulant $B_4=K+3$.
Infinite volume limit of kurtosis at the transition point
determined by the peak position of the susceptibility or the zero of
the skewness provides a diagnosis on the nature of transition as follows:
\begin{enumerate}
\item $\lim_{V\rightarrow\infty}K_X=-2$: first order,
\item $-2<\lim_{V\rightarrow\infty}K_X<0$: second order with the value
  determined by the universality class,
\item $\lim_{V\rightarrow\infty}K_X=0$: crossover.
\end{enumerate}
Infinite volume limit of the minimum value of the CLB cumulant is as follows:
\begin{enumerate}
\item
  $\lim_{V\rightarrow\infty}U_X
%  = \frac{2}{3} \frac{x^4+\Delta^4}{(x^2+\Delta^2)^2}$
  \neq 2/3$:
%  ($\Delta\neq0$):
  first or second order,
%\item
%  $\lim_{V\rightarrow\infty}U_X
%  \neq 2/3, \frac{2}{3}\frac{x^4+\Delta^4}{(x^2+\Delta^2)^2}$:
%  2nd order
\item $\lim_{V\rightarrow\infty}U_X=2/3$: crossover.
\end{enumerate}
%where $x=\langle\mathcal{O}\rangle$ and $2\Delta$ is the  inter-peak
%distance of double Gaussian distribution.
The reasoning for the first order phase transition case is given in
Appendix~\ref{sec:doublepeakdistribution} where the limit value of the
CLB cumulant is given in terms of the expectation value of $X$ in the
two phases.
%The discussion of the first
%order phase transition following the double Gaussian distribution for
%the CLB cumulant is given in
%Appendix~\ref{sec:doublepeakdistribution}.
Of course we do not {\it a priori} know these values which are
dictated by dynamics.  Therefore the limit value of the CLB cumulant
is not sufficient to distinguish between a first and a second order
transition.  The difference may become clear by looking at the volume
scaling.  For instance, if the volume scaling is given by an integer
power $V$, then the transition is considered as first order.

%% ukawa: there is a subtlety with CLB cumulant pointed out by Jin san.  I have checked his point.  He is correct in that the minimum of CLB cumulant deviates from the transition point defined by the peak susceptibility or skewness zero.  The minimum, however, does go to the transition point at V=infty, and CLB deviates away from 2/3 although the value differs from the one calculated at the peak height of susceptibility.  I have revised the Appendix A to incorporate this point.

\subsection{\label{sec:plaquette}Plaquette, gluon action density, and Polyakov loop}

The plaquette average is given by
\begin{equation}
  P=\frac{1}{18VN_{\text t}}
  \sum_{x,1\leq\nu<\rho\leq 4} \re W_{\nu\rho}^{1\times1},
\end{equation}
where the individual plaquette $W_{\nu\rho}^{1\times1}$ is defined in
Eq.~\eqref{eqn:plaquetteloop} and $V$ denotes the spatial lattice
volume $V=\Nxyz$.
The gauge action density is defined as
%\footnote{ This footnote will be
%  eliminated in the final version, but I show this for a purpose of
%  consistency check.  Outputs of the BQCD code are as follows,
%  \begin{align}
%    a &= \frac{1}{6VN_{\text t}} \sum_{x,\mu<\nu} \left( \frac{1}{3}
%      \re W_{\mu\nu}^{1\times1}(x) \right),
%    \\
%    b &= \frac{1}{6VN_{\text t}} \sum_{x,\mu<\nu} \left( \frac{1}{3}
%      \re W_{\mu\nu}^{1\times2}(x) \right),
%  \end{align}
%  for plaquette and rectangle part respectively.  Note that the sum in
%  $b$ there is a constraint for $\mu$ and $\nu$.  (JIN: I still don't
%  understand.  Are the other half, $W^{1\times2}_{\mu>\nu}$, not
%  summed?)  And then I construct the observable $G$ by
%  \begin{equation}
%    G=c_0(1-a)+2c_1(1-b).
%  \end{equation}
%}
\begin{equation}
  G
  =
  \frac{1}{6VN_{\text t}}
  \sum_x
  \left\{
    c_0
    \sum_{1\leq\nu<\rho\leq 4}
    \left(
      1-\frac{1}{3}
      \re  W_{\nu\rho}^{1\times1}(x)
    \right)
    +
    c_1
    \sum_{1\leq\nu,\rho\leq 4}
    \left(
      1-\frac{1}{3}
      \re  W_{\nu\rho}^{1\times2}(x)
    \right)
  \right\},
  \label{eqn:defG}
\end{equation}
and the Polyakov loop is defined by
\begin{equation}
  L=\frac{1}{3V}\sum_{{\bf x}} \tr
  \left[\prod_{x_4=1}^{N_{\text t}} U({\bf x},x_4,\nu=4)\right].
  \label{eqn:defpolyakovloop}
\end{equation}

For the three gluonic quantities defined above, writing $X=P$, $G$ or $L$,
the cumulants\footnote{
For the Polyakov loop susceptibility we define
$\chi_L= V\langle(L-\langle L\rangle)^2\rangle$
without a factor $N_{\text t}$.
}
are explicitly given by
%% ukawa: the factor N_t is included in the definition. ST I see.
\begin{align}
  \chi_X&= VN_{\text t}\langle(X-\langle X\rangle)^2\rangle \nonumber\\
  S_X &= \frac{
    \langle (X-\langle X\rangle)^3 \rangle
  }{
    \langle (X-\langle X\rangle)^2 \rangle^{3/2}
  } \\
  &= \frac{
    \langle X^3 \rangle -
    3\langle X^2 \rangle \langle X \rangle +
    2\langle X \rangle^3
  }{
    (\langle X^2\rangle - \langle X\rangle^2)^{3/2}
  }, \label{eqn:skewness} \\
  K_X &= \frac{
    \langle (X-\langle X\rangle)^4 \rangle
  }{
    \langle (X-\langle X\rangle)^2 \rangle^{2}
  } -3 \nonumber\\
  &= \frac{
    \langle X^4 \rangle -
    4 \langle X^3 \rangle \langle X \rangle -
    3 \langle X^2 \rangle^2 +
    12 \langle X^2 \rangle \langle X \rangle^2 -
    6 \langle X \rangle^4
  }{
    (\langle X^2\rangle - \langle X\rangle^2)^{2}
  }. \label{eqn:kurtosis}
\end{align}
Note that we include a factor $N_{\text t}$ in the susceptibility by convention.

\subsection{Fuzzy Polyakov loop}

The quantity $v^{(q)}$ defined in the winding expansion of the
determinant in Eq.~\eqref{eqn:Wdefwinding} is a sum of gauge loops
winding around the time direction $q$ times. In this sense they define
a fuzzy Polyakov loop.  For example, $v^{(1)}$ turns out to be a
normal Polyakov loop in the static limit up to an overall
normalization,
\begin{equation}
  v^{(1)}
  \stackrel{\kappa\rightarrow0}{=}
  -(2\kappa)^{N_{\text t}}2\cdot3 L,
  \label{eqn:staticlimitvL}
\end{equation}
where $L$ is the Polyakov loop in Eq.~\eqref{eqn:defpolyakovloop}.

\begin{figure}
  \subfloat[]{\includegraphics[scale=1]{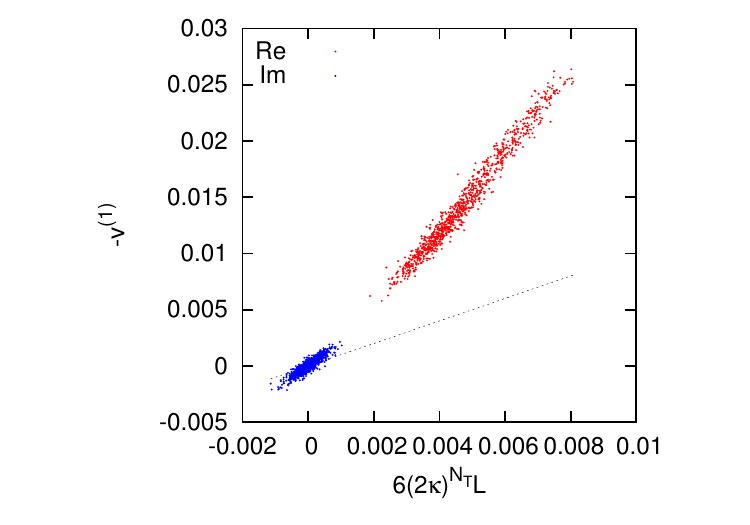}} \qquad
  \subfloat[]{\includegraphics[scale=1]{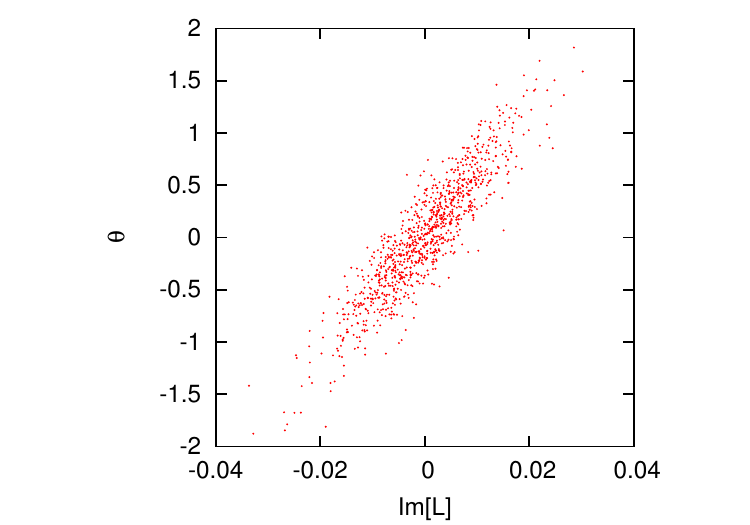}}
  \caption{\label{fig:poly_vqfp}
    (a) Correlation between the the fuzzy Polyakov loop $-v^{(1)}$ and
        Polyakov loop $L$ multiplied with $6(2\kappa)^{N_{\text t}}$
         on 1000 phase quenched configurations.
    Red and blue points respectively represent the real and imaginary part.
        The dotted black line shows the static limit for the fuzzy Polyakov loop
        given in Eq.~\eqref{eqn:staticlimitvL}.
    (b) correlation between the phase of determinant and the imaginary
    part of the Polyakov loop.
    The simulation parameters are as follows: $8^3\times4$, $\beta=1.60$,
    $\kappa=0.1371$ and $a\mu=0.205$.
}
\end{figure}

In Fig.~\ref{fig:poly_vqfp} (a), we show the correlation between
$-v^{(1)}$ and $6(2\kappa)^{N_{\text t}} L$.
% $8^3\times4$, $\beta=1.60$, $\kappa=0.1371$ and $a\mu=0.205$.
The real part as well as the imaginary part shows a strong correlation
in the parameter space where we investigate, albeit the deviation from
the static limit is significant.

As is seen from Eq.~\eqref{eqn:Wdefwinding}, the imaginary part of
$v^{(q)}$ contributes to the phase of the determinant.  Therefore a
correlation between the phase and the imaginary part of the Polyakov
loop is also expected.  It is indeed confirmed in
Fig.~\ref{fig:poly_vqfp}(b) where the phase is exactly computed from
$W(\mu/T)\in\mathbb{C}$ in Eq.~\eqref{eqn:WHHH} up to $2\pi$
periodicity.  Such a correlation was observed in
Ref.~\cite{deForcrand:1999cy} in the heavy mass region for the
staggered quark action.

%We measure a fuzzy Polyakov loop $v^{(1)}$ which is defined in the
%fermion determinant expressed by the winding
%expansion in Eq.~\eqref{eqn:Wdefwinding},
%\begin{equation}
%  \ln\det D(\mu)
%  =
%  \ln A_0
%  -
%  V\sum_{q\in\mathbb{Z}}
%  e^{q\mu/T}
%  v^{(q)}.
%  \label{eqn:windingexpansionln}
%\end{equation}
%where the lattice spatial volume $V$ is factored out in the second
%term.  We have computed $A_0$ and $v^{(q)}$ (In the paper
%\cite{Takeda:2011vd}, we denoted $V^{(q)}=Vv^{(q)}$) with
%$q=0,1,...,10$.

If the power of fugacity is promoted to an independent parameter for
each $q\in\mathbb{Z}$,
\begin{equation}
  e^{q\mu/T}\longrightarrow\lambda^{(q)},
\end{equation}
$v^{(q)}$ can be considered as the first derivative of the promoted
partition function $\mathcal{Z}_{\text{QCD}}$ in terms of the new
parameter,
\begin{equation}
  \langle
  v^{(q)}
  \rangle
  =
  -
  \frac{1}{N_{\text f}V}
  \left.\frac{\partial\ln \mathcal{Z}_{\text{QCD}}}{\partial\lambda^{(q)}}\right|_{\lambda^{(q)}=\exp(q\mu/T)},
\end{equation}
with
\begin{equation}
  \mathcal{Z}_{\text{QCD}}(..,\lambda^{(1)},\lambda^{(2)},\lambda^{(3)},...)
  =
  \int [\dd U]
  \exp\left\{-S_{\text G}[U]
    +
    N_{\text f}\ln A_0
    -
    N_{\text f}V\sum_{q\in\mathbb{Z}}
    \lambda^{(q)}
    v^{(q)}
  \right\}.
\end{equation}
In the end, we impose $\lambda^{(q)}=e^{q\mu/T}$ for all
$q\in\mathbb{Z}$ to restore the original theory.  Singularities of the
theory may be captured by this quantity.  Therefore we analyze higher
cumulants of $v^{(q)}$ defined by taking higher derivatives of $\ln
\mathcal{Z}_{\text{QCD}}$.  In practice, we exclusively analyze the
cumulants of $v^{(1)}$.

\subsection{Quark number}
The quark number density normalized by $T^3$ is given by
\begin{equation}
  \frac{n_{\text q}}{T^3}
  =
  \frac{1}{VT^3}
  \frac{\partial\ln\mathcal{Z}_{\text{QCD}}}{\partial(\mu/T)}
  =
  \frac{\langle Q_1\rangle}{VT^3}.
  \label{eqn:quarknumber}
\end{equation}
Following the general definition adopted in Sec.~\ref{sec:moments},
the other higher moments are given by
\begin{align}
  \frac{\chi_{\text q}}{T^2} &= \frac{1}{VT^3}
  % \frac{\partial^2}{\partial(\mu/T)^2}\ln\mathcal{Z}
  (\ln\mathcal{Z}_{\text{QCD}})^{(2)} = \frac{\langle Q_2\rangle-\langle
    Q_1\rangle^2}{VT^3},
  \\
  S_{\text q} &= \frac{(\ln\mathcal{Z}_{\text{QCD}})^{(3)}}
  {((\ln\mathcal{Z}_{\text{QCD}})^{(2)})^{3/2}} = \frac { \langle Q_3\rangle
    -3 \langle Q_2\rangle \langle Q_1\rangle + 2\langle Q_1\rangle^3}
  {(\langle Q_2\rangle-\langle Q_1\rangle^2)^{3/2}},
  \\
  K_{\text q} &= \frac{(\ln\mathcal{Z}_{\text{QCD}})^{(4)}}
  {((\ln\mathcal{Z}_{\text{QCD}})^{(2)})^{2}} = \frac { \langle Q_4\rangle -4
    \langle Q_3\rangle \langle Q_1\rangle -3 \langle Q_2\rangle^2 +12
    \langle Q_2\rangle \langle Q_1\rangle^2 -6 \langle Q_1\rangle^4 }
  {(\langle Q_2\rangle-\langle Q_1\rangle^2)^{2}},
\end{align}
where $(n)$ means the $n$-th derivative $\partial^n/\partial(\mu/T)^n$,
$Q_n (n=1,2,3,4)$ are given in Eq.~\eqref{eqn:Q}, and the CLB cumulant
takes the form,
\begin{equation}
  U_{\text q} = 1-\frac{ \langle Q_4 \rangle } {3 \langle Q_2
    \rangle^2 }.
  \label{eqn:CLBcumulantQ}
\end{equation}

\section{\label{sec:finitedensitymoment}Simulation
  results}

% \begin{itemize}
% \item higher moments results obs versus a$\mu$ plot
% \item comparison between ours and Kentucky $n-\mu$ plot
% \item comparison of with and without phase
% \item comment on mu-reweighting (ref next section for details) and
%   determine the peak of suscepbility and the minimum of the CLB
%   cumulant.
% \item volume scaling plot and discuss the order of transition.
% \end{itemize}

We now discuss simulation results for the expectation
value, susceptibility and higher cumulants.  In the figures we
only plot their real part since their imaginary part vanishes due to
symmetry.
% The same goes for other physical quantities given later.

\subsection{\label{sec:simul-results-from-mu}Numerical evaluation of $\mu$-reweighting}
%In this section, we evaluate $\mu$-reweighting by using actual simulation data.
%The details of simulation and the definition of physical quantities
%will be given in section \ref{sec:simulationparameters}
%and \ref{sec:defquantities} respectively.

% \begin{itemize}
% \item comparison of Three expansion schemes. Log Taylor is the best
%   and in the following we show only this expansion.  In the
%   previouse section, all the $\mu$-reweighting results are Log
%   Taylor.
% \item Various Volumes. Volume dependence of the convergence range is
%   small.
% \item different reweighting points
% \end{itemize}

\begin{figure}
  \subfloat[]{\includegraphics{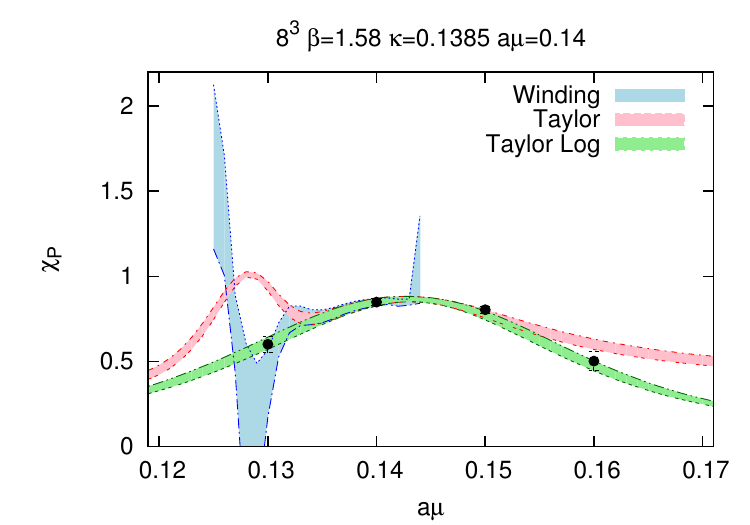}} \qquad
  \subfloat[]{\includegraphics{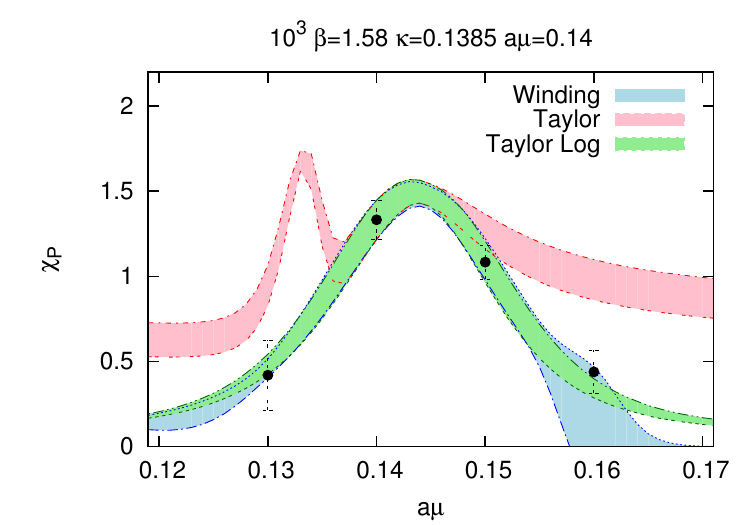}}
  \caption{\label{fig:mureweighting_expansion_comparison}Comparison of
    three expansion schemes of $\mu$-reweighting for the
    susceptibility of plaquette.  Winding expansion: blue band; Taylor
    expansion: red band; Taylor expansion of the logarithm of the
    determinant:green band.  Black symbols are direct simulation data.
    The original $\mu$-value is $a\mu=0.14$. (a) $8^3\times4$, (b)
    $10^3\times4$.}
\end{figure}

In Fig.~\ref{fig:mureweighting_expansion_comparison}, we compare the
three expansion schemes introduced in Sec.~\ref{sec:mureweighting},
taking the susceptibility of plaquette for illustration.  The starting
value is $a\mu=0.14$, and the results of $\mu$-reweighting are shown
by the one standard deviation error bands.  The simulation paramaters
are given in the figure.  The performance of $\mu$-reweighting can be
measured by comparison of the bands with actual measurements away from
$a\mu=0.14$ plotted by filled circles.  Comparing the results for
$8^3\times 4$ lattice in (a) and for $10^3\times 4$ lattice in (b), we
see that the winding expansion works better for larger volume.  The
Taylor expansion develops a fake transition around $a\mu=0.128$ on
$8^3\times4$ lattice and around $a\mu=0.133$ on $10^3\times4$ lattice,
respectively.
% in addition to the real
%transition around $a\mu=0.14$.  It is likely that the truncation error
%of this expansion induces such a fake transition.
% For larger volume, the fake transition point is getting close to the
% real transition point.
The applicable range of $\mu$-reweighting for this expansion becomes
smaller for larger volumes.
%A possible explanation of this behavior is that the
%coefficient of the Taylor expansion is proportional to the volume
%therefore the convergence becomes worse for larger volume.
In contrast to the two expansions, the Taylor expansion of the
logarithm is working well for both lattice sizes and the applicable
range is quite wide compared with the other expansion schemes.

\begin{figure}
  \includegraphics[scale=1.2]{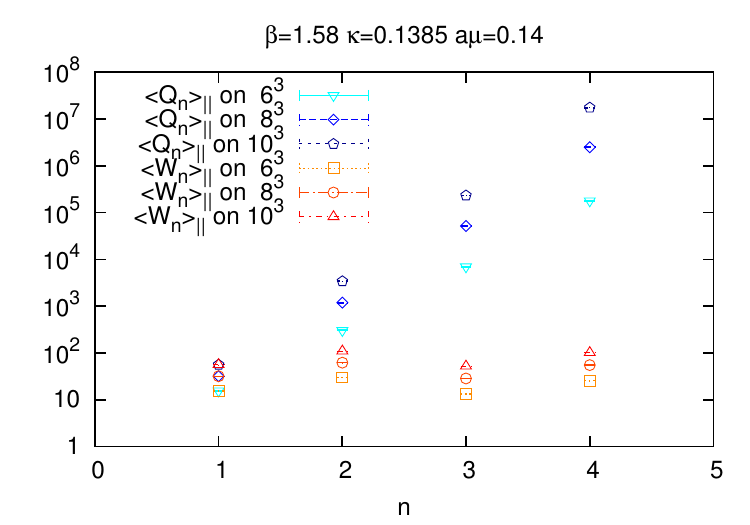}
  \caption{\label{fig:coefficienthierarchy}Phase quenched average of
    coefficients $Q_n$ and $W_n$ as a function of $n$.  $Q_n$ are for
    the Taylor expansion of the determinant, and $W_n$ for the Taylor
    expansion of the logarithm of the determinant.  The spatial volume
    is changed from $6^3$ to $10^3$ while the temporal lattice size is
    fixed to $N_{\text t}=4$.  Error bars are too small to see at this
    scale.}
\end{figure}

A possible explanation of this behavior is as follows.  As is seen
from Eq.\eqref{eqn:W}, the coefficients of Taylor expansion of the
logarithm $W_n$ are made of single trace whose magnitude would be
proportional to the reduced space, namely the spatial lattice size
$W_n\propto V$.  Since this holds for all $n=1,2,3,4,...$, the
magnitude of $W_n$ would not increase for larger $n$.  Such a tendency
is observed in $\langle W_n \rangle_{||}$ as shown in
Fig.~\ref{fig:coefficienthierarchy}.  On the other hand, the
coefficients of Taylor expansion $Q_n$ are made from a product of
$W_n$.  Hence the dominant volume scaling is expected to be
$Q_n\propto V^n$, and this tendency is seen in
Fig.~\ref{fig:coefficienthierarchy}.
%This consideration explains why the Taylor expansion of the logarithm is better than Taylor expansion.
In this way, we conclude that the Taylor expansion of the logarithm of
the determinant is the best among our choices.  This expansion scheme
is used in the following $\mu$-reweighting results.

% Actually, the size of error bar at the peak is slightly smaller than
% that at the original simulated $\mu$ point. Why?

\begin{figure}
  \includegraphics[scale=1.2]{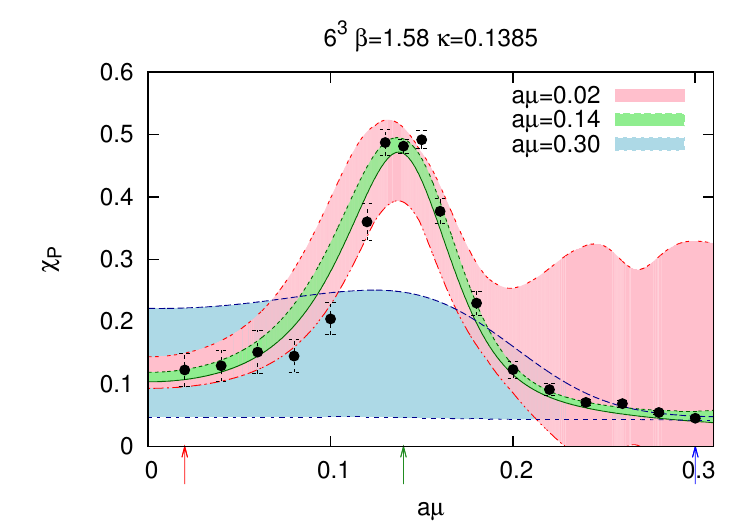}
  \caption{\label{fig:mureweighting_mupoint_comparison}Comparison of
    the susceptibility of plaquette calculated from $\mu$-reweighting
    and from direct simulation.  Black symbols show results from
    direct simulations.  Colored regions show one standard deviation
    bands of $\mu$-reweighted results.  Three different ensembles are
    used for $\mu$-reweighting, and their respective simulation
    points, $a\mu$, are labeled in the figure and also pointed at the
    horizontal axis by the same colored arrow}
\end{figure}

Lastly, we compare $\mu$-reweighting from ensembles at three original
values of $\mu$ given by $a\mu=0.02$ and $0.14$ and $0.30$ in
Fig.~\ref{fig:mureweighting_mupoint_comparison}.  The statistics for
each ensemble are roughly the same order.  We observe that the data
reweighted from $a\mu=0.14$ shows an excellent agreement with the
actual simulation data plotted by filled circles over a wide range
from $a\mu=0.02$ to $0.30$.  Also the estimated errors do not change
much over this region.  On the other hand, the reweighting from
$a\mu=0.02$ and $0.30$ do not work well away from the original value.
This may mean that not only the truncation error of the expansion but
also the overlap issue is very important.  The configurations
generated at $a\mu=0.14$ are sampled from both low density phase and
high density phase.  Therefore the distribution of the plaquette has
large overlaps with both phases.  On the other hand, the
configurations generated at $a\mu=0.02$ are mainly sampled from the
low density phase, and hence the overlap with the high density phase
region is very small.  An opposite situation holds for the
configurations generated at $a\mu=0.30$.
%In the end,
%$\mu$-reweighting from around transition region is very excellent since in
%such a region configurations in both phases are equally sampled.

%%ukawa: I feel that the following sentence is not necessary. ST OK
%We remark that
%we do not use any noise estimator to evaluate the trace in the
%coefficients of the expasion but they are computed exactly instead.
%Although this is very expensive calculation, we are free from the
%noise contamination.

\subsection{Phase-reweighting factor}
\begin{figure}
  \subfloat[]{\includegraphics{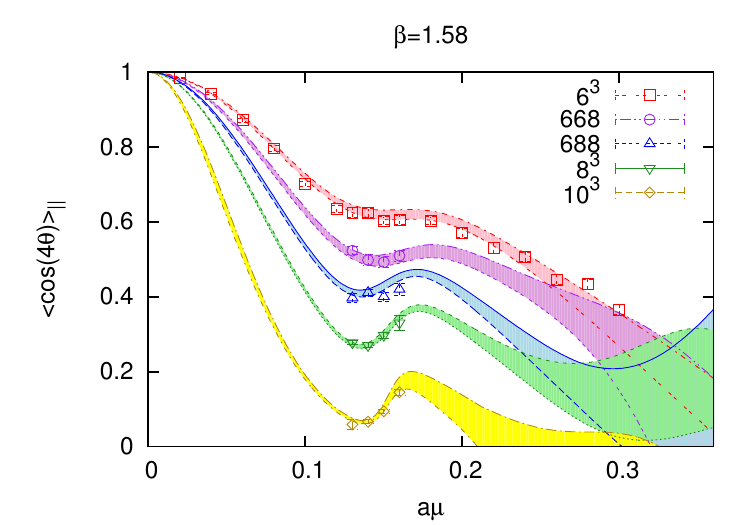}} \qquad
  \subfloat[]{\includegraphics{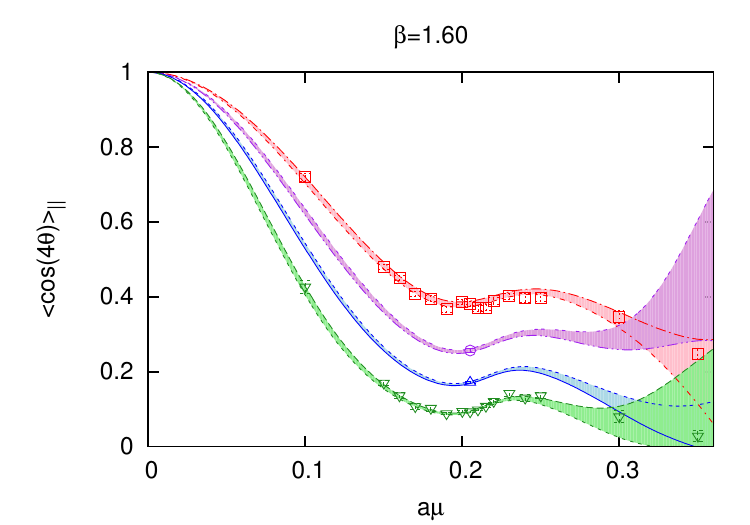}}
  \caption{\label{fig:phasereweighting}The phase-reweighting factor as
    a function of chemical potential at (a) $\beta=1.58$ and (b) $\beta=1.60$.
    % The baryon chemical potential in temperature unit is shown in the upper x-axis.
    The spatial volume is changed from $6^3$ to $10^3$ ($6^3$ to $8^3$)
    for $\beta=1.58$ ($\beta=1.60$).  Filled curves show $1\sigma$
    error band of the $\mu$-reweighed data from the original point at
    $a\mu=0.14$ and $0.205$ for $\beta=1.58$ and $1.60$ respectively.
%    Some details about the $\mu$-reweighting will be discussed in
%    Sec.~\ref{sec:reweighting}.
  }
\end{figure}

%Before showing results of moments, first of all we show the
%phase-reweighting factor to see a situation of the sign problem.
In Fig.~\ref{fig:phasereweighting} we show the phase-quenched average of
the phase-reweighting factor as a function of $a\mu$ at $\beta=1.58$ and
$1.60$.  The $\mu$-reweighting one standard deviation error bands
from $a\mu=0.14$ at $\beta=1.58$ and from $a\mu=0.205$ at $\beta=1.60$
are also shown.
% whose
%details will be given in Sec.~\ref{sec:reweighting}
For larger volumes, the reweighting factor tends closer to zero, such
that the sign problem becomes more serious as expected.  However,
since the phase-reweighting factor remains non-vanishing beyond
statistical errors, the sign problem is under control for the lattice
volumes and the parameter sets used in the present simulations.

%%(ukawa: I am not sure if this argument is really meaningful. ST:In section J, it is discussed in some details. This is just an observation.)
An interesting observation is that there is a local minimum around
$a\mu=0.14$ ($a\mu=0.2$) for $\beta=1.58$ ($\beta=1.60$).  This is
related to a change in the partition function, which usually appears
as a consequence of a phase transition.  It will be apparent when we
discuss the behavior of the pressure in Sec.~\ref{sec:pressure}.

%An interesting observation is that there is a local minimum around
%$a\mu=0.14$ ($a\mu=0.2$) for $\beta=1.58$ ($\beta=1.60$).  We expect
%this to be the region of transition (or rapid crossover), which can be
%understood as follows.  Remembering the definition of the reweighting
%factor $\langle e^{i4\theta}\rangle_{||}=\mathcal{Z}_{\text{QCD}}/\mathcal{Z}_{\text{QCD}_{||}}$, we observe that a zero of the average
%reweighting factor is related with a zero of the QCD partition
%function, which is a transition point of the theory.  Of course, we do
%not expect a real phase transition but only a remnant for finite
%volumes, and hence no actual zeros for the real parameter space but
%just a remnant, such as a minimum.  The dip in
%Fig.\ref{fig:phasereweighting} may be considered to be such a remnant,
%in which case the parameter region around $a\mu=0.14$ ($a\mu=0.2$) is
%expected to contain the transition point for $\beta=1.58$
%($\beta=1.60$).  \jinnote{clear this up}

\subsection{Comparison between QCD and phase-quenched QCD}
\begin{figure}
  \subfloat[]{\includegraphics{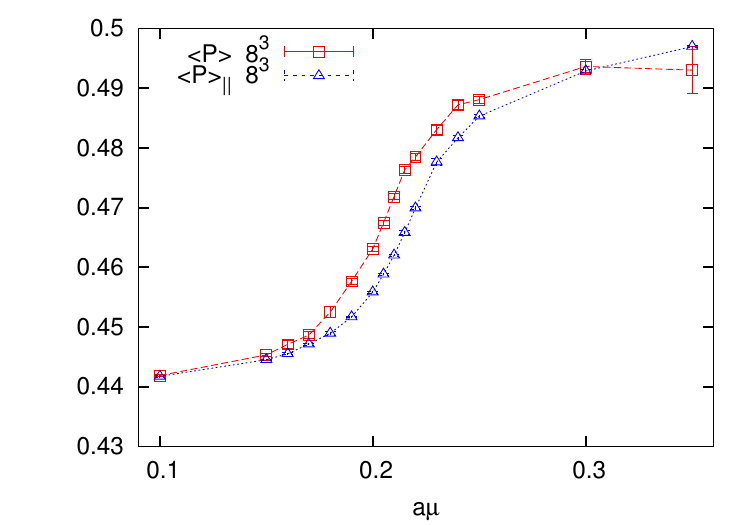}} \qquad
  \subfloat[]{\includegraphics{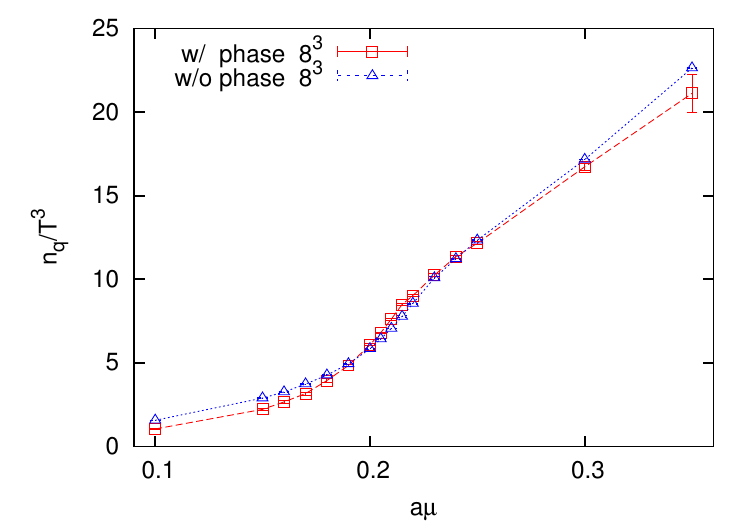}}
  \caption{\label{fig:phasecomparison}(a) Plaquette average and (b)
    quark number density as a function of $a\mu$ at $\beta=1.60$ on a
    $8^3\times4$ lattice.  Red squares are results for full QCD and
    blue triangles for phase-quenched QCD.}
\end{figure}

%%(ukawa: I am not sure if this observation is important. ST: I heard some people are interested in this kind of argument, say Hanada-san. This is may be useful information for them and AdS/CFT community.)
%Let us see an effect of including the complex phase.
Fig.~\ref{fig:phasecomparison} compares the average value of
plaquette and the quark number density calculated with and without the
phase of the quark determinant at $\beta=1.60$ on a $8^3\times4$
lattice.  Apart from a small difference resembling a shift in $a\mu$
in the region of rapidly increasing plaquette, the effect of inclusion
of the phase is quite small in the figure for large values of $a\mu$.
Such a trend is observed also for higher moments and other physical
quantities.  Similar observation has been reported in
Ref.~\cite{Sasai:2003py} in $N_{\text f}=2$ QCD by the phase
reweighting method.  In Ref.~\cite{Hanada:2012es} it was argued that
such a phenomena should hold at the parameter points outside of the
charged pion condensation phase in the large $N_c$ limit.

%%ukawa: I do not think the following sentence is necessary because phase-included results are the correct results.  Phase-quenched results are not physical results. ST I agree with you to omit the sentence.
%As we have seen, the effect of the inclusion of the phase is tiny,
%therefore
%in the following, we do not show phase-quenched results, and
%exclusively show only phase-included results.

\subsection{Comparision with the Kentucky group}
\begin{figure}
  \subfloat[]{\includegraphics[height=1.8in]{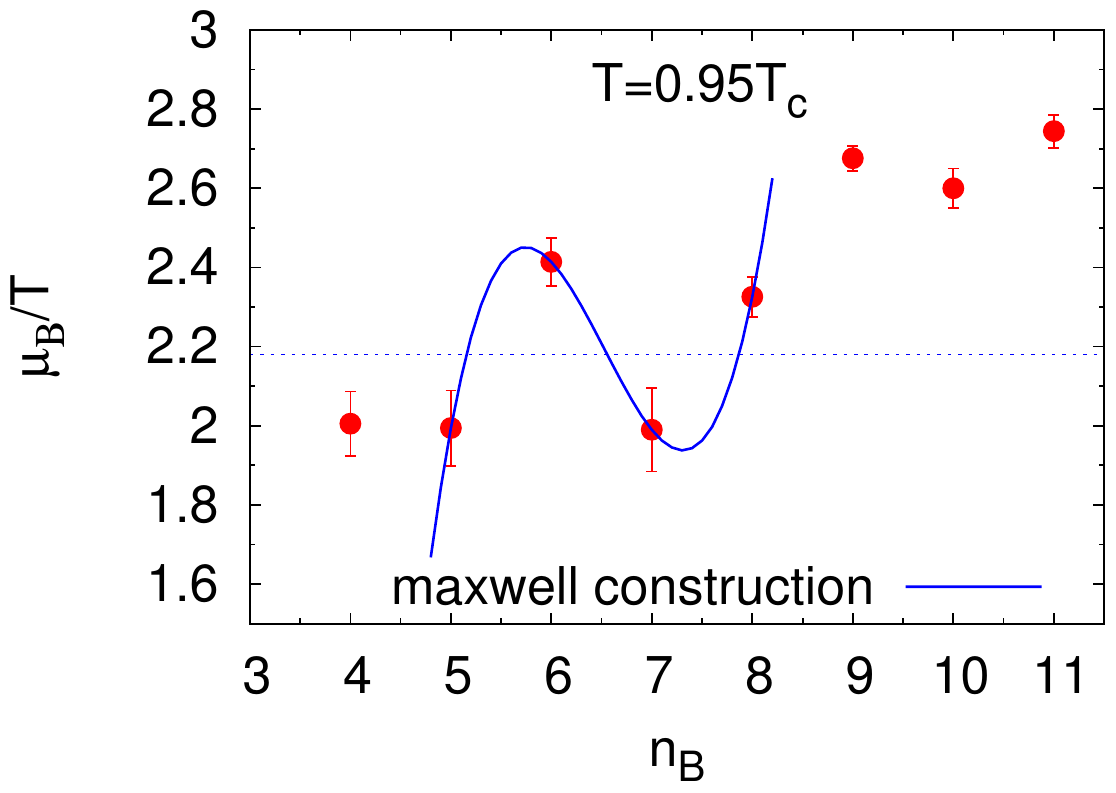}}
  \qquad
  \subfloat[]{\includegraphics[height=1.8in]{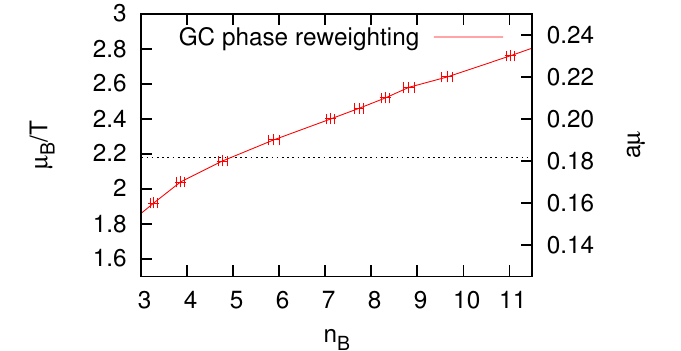}}
  \caption{\label{fig:Kentuckycomparison}Comparison of canonical
    (filled circles in (a)) and grand canonical (open symbols in (b))
    results for the relation between baryon chemical potential
    $\mu_{\text B}/T$ and baryon number $n_{\text B}$.  Canonical
    results are from Ref.\cite{Li:2010qf} by the Kentucky group.
    Simulation parameters are $\beta=1.60$, $\kappa=0.1371$ on a
    lattice of size $6^3\times4$.}
\end{figure}

%%ukawa: Reproducing Kentucky figure without permission from APS is probably copy right violation.  Can you read off the numerical values from their figure and plot them on (b)? ST I could not find the value. APS may contact us if something is wrong.
%Let us see a comparision between the grand canonical and canonical
%approach.
The Kentucky group~\cite{Li:2010qf} carried out a canonical simulation
at $\beta=1.60$ and $\kappa=0.1371$ on a $6^3\times 4$ lattice
employing the same gluon and quark actions as in the present study.
In their work, the quark or baryon chemical potential $\mu_{\text
  q}=\mu_{\text B}/3$ is measured at fixed quark or baryon number
$n_{\text q}=3n_{\text B}$, and they constructed an S-shape in their
baryon number versus baryon chemical potential plot.  In our grand
canonical simulation, on the other hand, the input is the chemical
potential and the output is the quark number.  We numerically compare
the two approaches in Fig.~\ref{fig:Kentuckycomparison} for the same
parameter set; filled symbols in (a) with vertical error bars are the
canonical results from Fig.~7 (bottom) in Ref.~\cite{Li:2010qf},
whereas open symbols in (b) with horizontal error bars are our grand
canonical results.
%JIN: changed slightly for the current figures

Outside the transition region, say $n_{\text B}\le4$ and $n_{\text B}\ge10$,
results from the two approaches agree with each other.
%Therefore our calculation is consistent with that of Kentucky group.
However, the two approaches show completely different behavior around
the transition region.  Graphically speaking in
Fig.~\ref{fig:Kentuckycomparison}, while the canonical results can be
made to produce an S-shape presumed from a first order transition, the
grand canonical results are expected to show a smooth behavior and
examination of higher cumulants such as susceptibility is required for
an indication of a transition.  The results of cumulant analyses,
however, suggest a numerical difference: the Maxwell construction of
the canonical results implies $\mu_{ \text B}/T\approx 2.2$ at the
transition, whereas the peak of quark number susceptibility from grand
canonical results in this study takes place at $\mu_{\text B}/T\approx
2.5$.  In principle the two approaches should lead to similar results
if the infinite volume limit is taken carefully.

%JIN: changed slightly for the current figures

\subsection{\label{sec:susceptibility}Susceptibility}

\begin{figure}
  \includegraphics[scale=1.2]{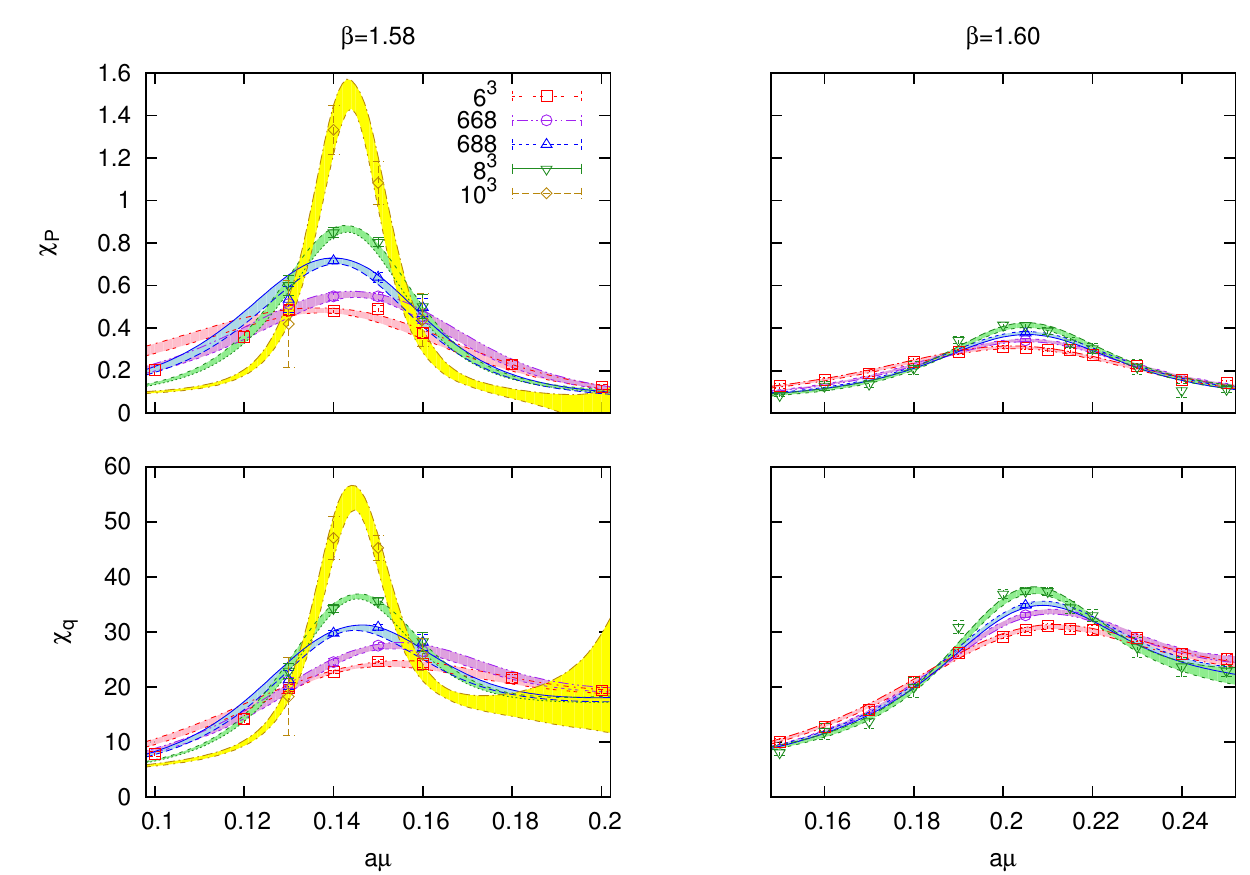}
  \caption{\label{fig:susceptibility}Susceptibility of plaquette
    (upper) $\chi_P$ and quark number density (lower) $\chi_{\text q}$
    as functions of $a\mu$ at $\beta=1.58$ (left) and $\beta=1.60$
    (right) for various spatial volumes.  }
\end{figure}

The susceptibility of plaquette $\chi_P$ and quark number density
$\chi_{\text q}/T^2$ are shown in Fig~\ref{fig:susceptibility}.  We
plot not only the actual simulation data with error bars but also the
one standard deviation $\mu$-reweighting band.  We observe a clear
volume dependence at $\beta=1.58$; the peak grows rapidly for larger
volume.  At $\beta=1.60$ the peak still grows with volume but the rate
is much milder.  The susceptibilities for gauge action density,
Polyakov and fuzzy Polyakov loop also show similar tendency.
Therefore it is likely that there is a phase transition at
$\beta=1.58$ while the situation at $\beta=1.60$ requires further
quantitative analyses.

\begin{figure}
  \subfloat[]{\includegraphics{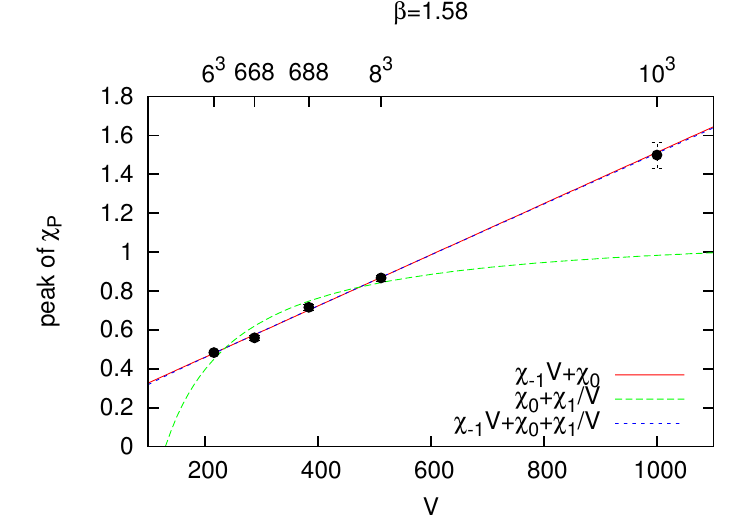}} \qquad
  \subfloat[]{\includegraphics{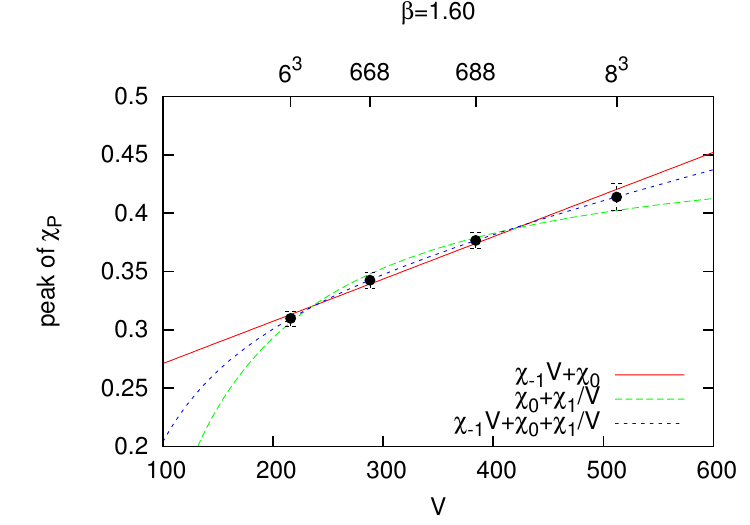}} \\
  \subfloat[]{\includegraphics{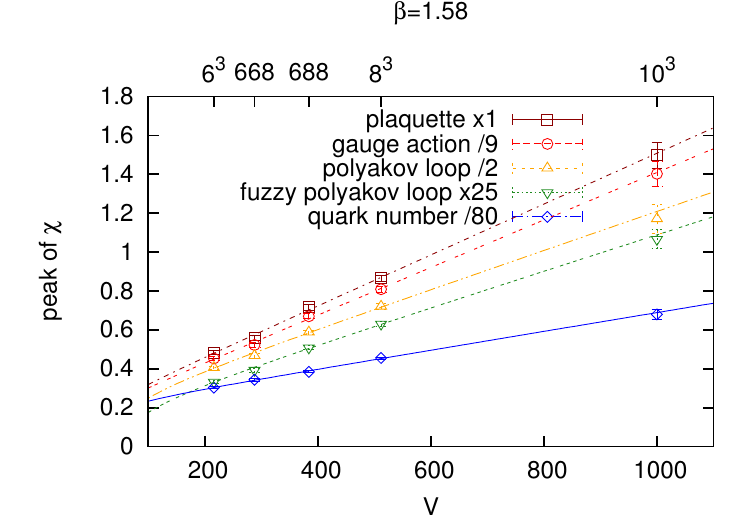}} \qquad
  \subfloat[]{\includegraphics{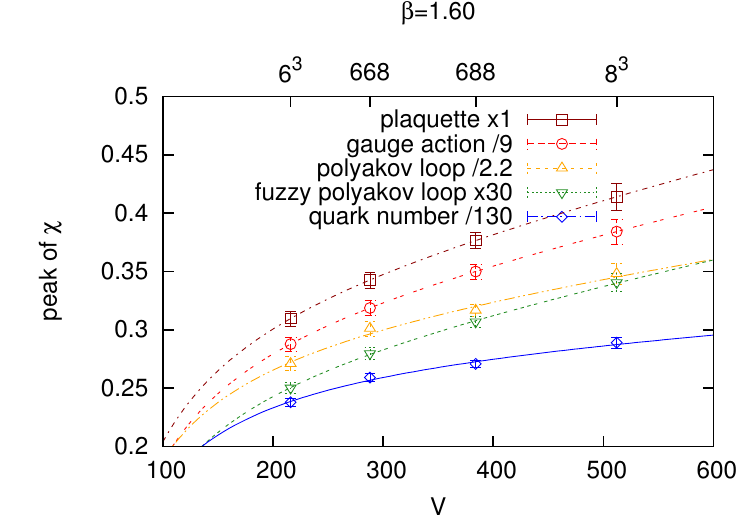}}
  \caption{\label{fig:susceptibility2}Upper panels show volume scaling
    of the peak value of $\chi_P$ for (a) $\beta=1.58$ and (b) $1.60$
    together with three types of fits.  Lower panels show volume
    scaling plots for all observables together with the fitting form
    S3 defined in the text.  Vertical scales are adjusted. }
\end{figure}

%Let us now address the question of the order of the phase transition.
We plot in Fig.~\ref{fig:susceptibility2} the volume dependence of the
peak height of $\chi_P$ for (a) $\beta=1.58$ and (b) $\beta=1.60$.
The peak position and the maximum value of $\chi_P$ is determined by
the $\mu$-reweighting.  The result for $\beta=1.58$ shows a clear
linear volume dependence, while that for $\beta=1.60$ is rather weak.

To draw a quantitative conclusion, we first try a fitting of data with
the functional form
\begin{equation}
  \chi_P^{\max}=aV^b+c,
\end{equation}
where $a$, $b$ and $c$ are fitting parameters.  It turns out that for
$\beta=1.58$ the exponent $b$ is consistent with 1 with a reasonable
error bar and reduced $\chi^2$.  On the other hand, the fit for
$\beta=1.60$ is very unstable and it is difficult to obtain a
meaningful exponent.  In the following, we assume a volume dependence
with integer powers of $V$ of the form
\begin{equation}
  \chi_P^{\max}=\chi_{-1}V+\chi_0+\chi_1/V,
  \label{eqn:susvolumescaling}
\end{equation}
and consider three cases,
\begin{enumerate}
  \item[S1] setting $\chi_{1}=0$
  \item[S2] setting $\chi_{-1}=0$
  \item[S3] no constraint
\end{enumerate}
%%ukawa:I think it is better to directly discuss results since it will be easier to understand. ST: Yes, your argument is better.
%If the fitting form 1 describes the date well
%or the resulting fit parameter $\chi_{-1}$ in the fitting form 3 is significantly
%different from zero then the 1st order phase transition is suggested.
%On the other hand, if the $\chi_{-1}$ in the fitting form 1 is consistent with zero or
%the fitting form 2 describes the data well then the crossover is indicated.
The results of the fits are summarized in Table~\ref{tab:chi_b1.58}
for $\beta=1.58$ and in Table~\ref{tab:chi_b1.60} for $\beta=1.60$ for
all susceptibilities we consider.  In the bottom panels (c) and (d) in
Fig.~\ref{fig:susceptibility2}, the volume scaling behavior for all
physical quantities are shown together with the fitting form S3.

Let us first look at Table~\ref{tab:chi_b1.58}.  For all five
observables, the fitting form S1 exhibits a reasonable reduced
$\chi^2$, and the coefficient $\chi_{-1}$ is well determined and
non-zero with less than a percent error.  This situation holds even if
one adds a $1/V$ term (fitting form S3), with the parameters
$\chi_{-1}$ and $\chi_0$ keeping values consistent with those from the
fitting form S1.  In a sharp contrast, dropping the term linear in $V$
(fitting form S2) leads to an unacceptably large reduced $\chi^2$.  We
conclude that there is a first order phase transition at $\beta=1.58$.

At $\beta=1.60$ in Table~\ref{tab:chi_b1.60}, the fitting form S1 also
provides a reasonable fit for all observables with a non-zero
$\chi_{-1}$ at a 10\% error level.  However, the fitting form S2
without the term linear in volume also yields fits of similar quality.
While a large negative coefficient $\chi_1$ of the $1/V$ term in the
latter fit does not seem natural, we are not able to exclude such a
possibility on other grounds.  With present data alone, it is
difficult to draw a clear distinction between a weak but first order
phase transition and a crossover at $\beta=1.60$.  Data for a larger
spatial lattice volume, \textit{e.g.}, $10^3$, will help, but it seems
very hard to accumulate enough statistics; the average of the fermion
phase is already rather small for our largest spatial volume of $8^3$
(see Fig.~\ref{fig:phasereweighting}).

\begin{table}
  \caption{\label{tab:chi_b1.58}Fitted values of parameters and
    $\chi^2/{\text{dof}}$ in the volume scaling form of susceptibility in
    Eq.~\eqref{eqn:susvolumescaling} for $\beta=1.58$.
    % For the gauge action case, the shown value is understood as %$\chi_{\text shown here}=\chi_{\text actual}N_{\text t}$.
    Values without errors are fixed during the fit.}
  \begin{ruledtabular}
    \begin{tabular}{l|c|l|l|l|l}
      observable          & fitting form & $\chi_{-1}$     & $\chi_{0}$    & $\chi_{1}$    & $\chi^2/{\text{dof}}$ \\
      \hline
                          & S1 & $0.001318(53)$  & $0.195(19)$   & $0$           & $0.853$              \\
      plaquette           & S2 & $0$             & $1.130(22)$   & $-147.1(6.4)$ & $31.5$               \\
                          & S3 & $0.00130(14)$   & $0.206(99)$   & $-2(16)$      & $1.27$               \\
      \hline
                          & S1 & $0.01106(44)$   & $1.65(16)$    & $0$           & $0.878$              \\
      gauge action        & S2 & $0$             & $9.49(19)$    & $-1231(53)$   & $31.4$               \\
                          & S3 & $0.0110(11)$    & $1.72(82)$    & $-10 (138)$   & $1.31$               \\
      \hline
                          & S1 & $0.002111(94)$  & $0.353(33)$   & $0$           & $0.648$              \\
      Polyakov loop       & S2 & $0$             & $1.816(38)$   & $-224(11)$    & $17.8$               \\
                          & S3 & $0.00199(28)$   & $0.44(19)$    & $-14(31)$     & $0.866$              \\
      \hline
                          & S1 & $0.0000400(14)$ & $0.00466(48)$ & $0$           & $0.997$              \\
      fuzzy Polyakov loop & S2 & $0$             & $0.03258(59)$ & $-4.29(16)$   & $30.3$               \\
                          & S3 & $0.0000370(39)$ & $0.0069(28)$  & $-0.36(44)$   & $1.18$               \\
      \hline
                          & S1 & $0.0399(19)$    & $15.74(74)$   & $0$           & $0.488$              \\
      quark number        & S2 & $0$             & $45.04(75)$   & $-4785(241)$  & $23.5$               \\
                          & S3 & $0.0384(46)$    & $17.0(3.5)$   & $-216(600)$   & $0.668$              \\
    \end{tabular}
  \end{ruledtabular}
\end{table}

\begin{table}
  \caption{\label{tab:chi_b1.60}Fitted values of parameters and
    $\chi^2/{\text{dof}}$ in the volume scaling form of susceptibility
    for $\beta=1.60$.}
  \begin{ruledtabular}
    \begin{tabular}{l|c|l|l|l|l}
      observable          & fitting form & $\chi_{-1}$      & $\chi_{0}$    & $\chi_{1}$   & $\chi^2/{\text{dof}}$ \\
      \hline
                          & S1 & $0.000362(39)$   & $0.235(13)$   & $0$          & $0.497$              \\
      plaquette           & S2 & $0$              & $0.472(14)$   & $-35.8(3.9)$ & $1.02$               \\
                          & S3 & $0.00022(15)$    & $0.332(99)$   & $-15(15)$    & $0.00052$            \\
      \hline
                          & S1 & $0.00302(33)$    & $1.97(11)$    & $0$          & $0.546$              \\
      gauge action        & S2 & $0$              & $3.95(12)$    & $-299(33)$   & $0.924$              \\
                          & S3 & $0.0017(13)$     & $2.83(82)$    & $-132( 127)$ & $7.32\times10^{-7}$  \\
      \hline
                          & S1 & $0.000555(72)$   & $0.486(25)$   & $0$          & $1.03$               \\
      Polyakov loop       & S2 & $0$              & $0.855(25)$   & $-56.4(7.3)$ & $1.07$               \\
                          & S3 & $0.00029(28)$    & $0.67(18)$    & $-28(28)$    & $1.08$               \\
      \hline
                          & S1 & $0.00001028(95)$ & $0.00624(32)$ & $0$          & $0.714$              \\
      fuzzy Polyakov loop & S2 & $0$              & $0.01303(33)$ & $-1.034(96)$ & $1.47$               \\
                          & S3 & $0.0000062(36)$  & $0.0090(24)$  & $-0.43(37)$  & $0.0651$             \\
      \hline
                          & S1 & $0.0222(23)$     & $26.63(79)$   & $0$          & $2.12$               \\
      quark number        & S2 & $0$              & $41.49(81)$   & $-2286(235)$ & $0.801$              \\
                          & S3 & $0.0067(89)$     & $37.1(5.9)$   & $-1625(909)$ & $1.03$               \\
    \end{tabular}
  \end{ruledtabular}
\end{table}

\subsection{Skewness}
\begin{figure}
  \includegraphics[scale=1.2]{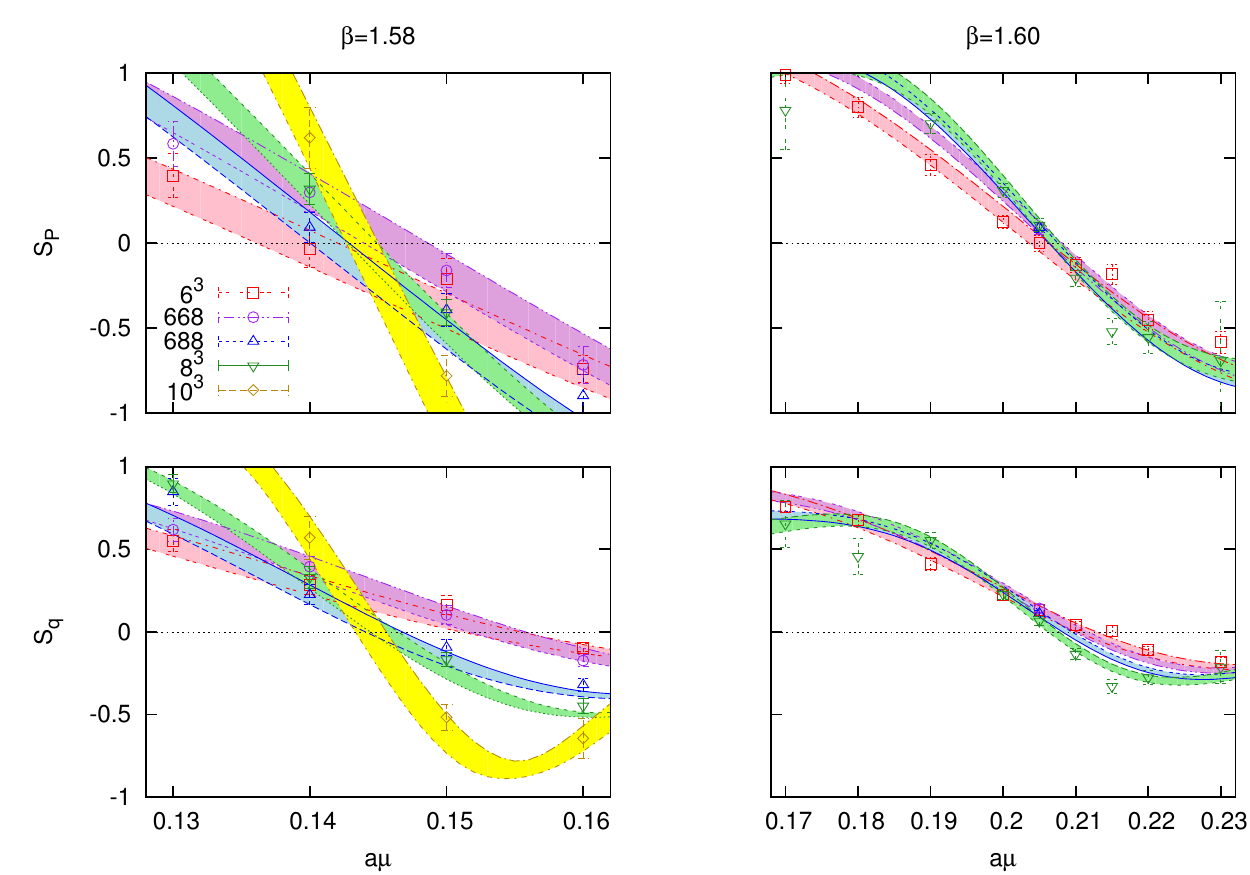}
  \caption{\label{fig:skewness}Skewness of plaquette $S_P$ (upper
    panels), and quark number density $S_{\text q}$ (lower panels) as
    functions of $a\mu$ at $\beta=1.58$ (left) and $\beta=1.60$
    (right).}
\end{figure}

The skewness of plaquette and quark number density are shown in
Fig.~\ref{fig:skewness}.  The zero of the skewness yields an estimate
of the transition point and the slope at the zero is expected to
negatively increase with volume. The latter feature is apparent in
Fig.~\ref{fig:skewness}.  The zeros estimated by $\mu$-reweighting are
consistent with the peak position of the susceptibility for each
observable and volume.  We find the volume dependence of the position
of zero to be less than 10\%.

%%%%%%%%%
%%%%%%%%%

\subsection{Kurtosis}

\begin{figure}
  \includegraphics[scale=1.2]{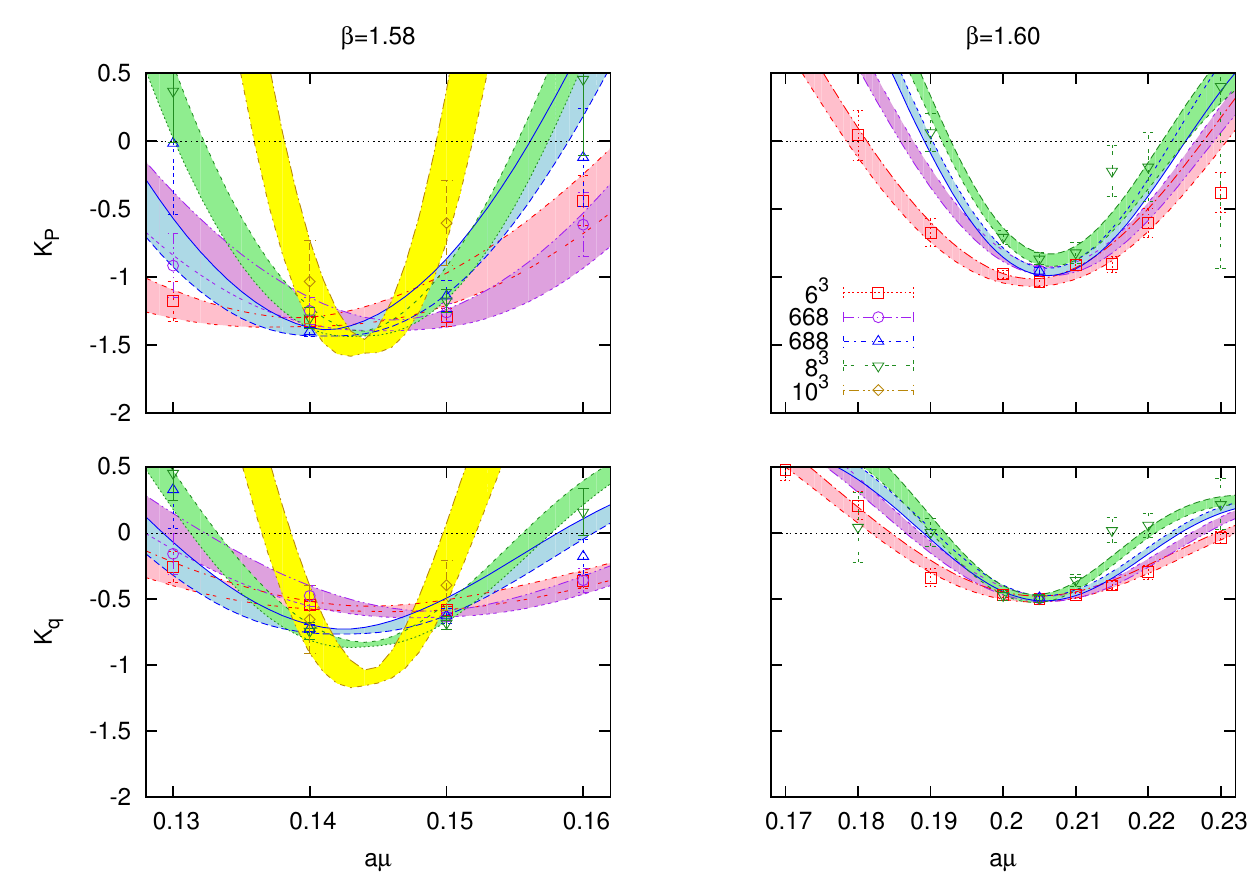}
  \caption{\label{fig:kurtosis}Kurtosis of plaquette $K_P$ (upper
    panels) and quark number density $K_{\text q}$ (lower panels) as
    functions of $a\mu$ at $\beta=1.58$ (left) and $\beta=1.60$
    (right).}
\end{figure}

\begin{figure}
  \subfloat[]{\includegraphics{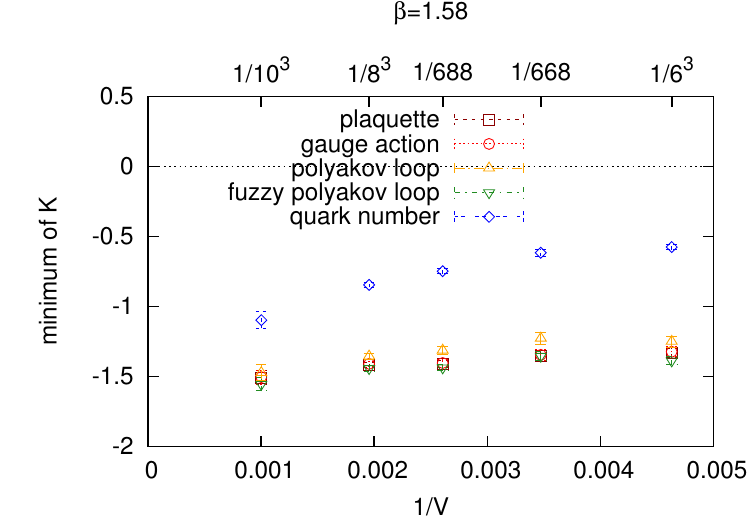}} \qquad
  \subfloat[]{\includegraphics{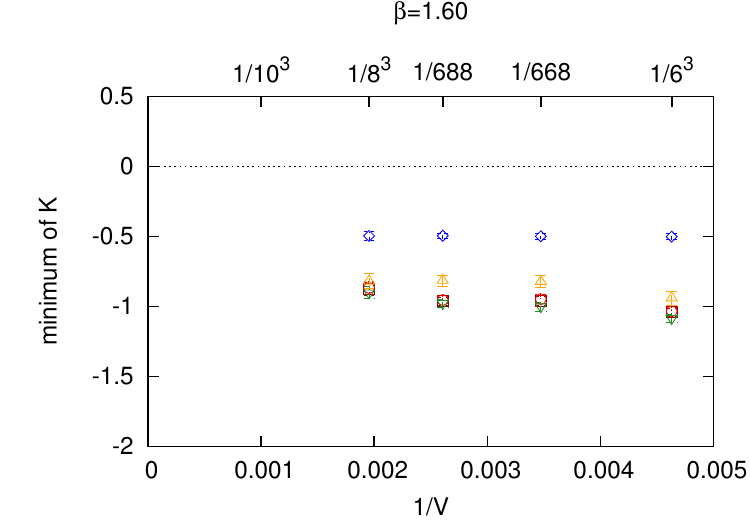}}
  \caption{\label{fig:kurtosis2}Volume scaling of the minimum value of
    the kurtosis for all physical quantities at (a) $\beta=1.58$ and
    (b) $\beta=1.60$.}
\end{figure}

The results of the kurtosis of plaquette and quark number density are
plotted in Fig.~\ref{fig:kurtosis}.  We observe a dip which becomes
sharper for larger volumes.  We also find that the peak position of
the susceptibility and the position of the minimum of the kurtosis is
consistent with each other for all physical quantities and each
volume.  These features are as expected from a simple double Gaussian
model discussed in Appendix~\ref{sec:doublepeakdistribution}.

Fig.~\ref{fig:kurtosis2} shows volume scaling of the minimum of
kurtosis for all observables.  At $\beta=1.58$, the minimum decreases
for larger volumes.  Infinite volume extrapolations assuming
polynomials in $1/V$, however, do not yield values close to $-2$
expected for a first order phase transition.  For $\beta=1.60$, the
minimum shows only weak volume dependence, and even increases slightly
for larger volumes.

Since kurtosis is composed of the fourth order cumulants, statistical errors
are significantly larger compared to the second order cumulants
(compare Fig.~\ref{fig:susceptibility} and Fig.~\ref{fig:kurtosis}).
Furthermore, the curvature at the minimum is expected to increase
quadratically in $V$. Unless data at the original value is precise,
$\mu$-reweighting may find hard time estimating the bottom of a sharp valley.
We feel that these features make kurtosis a rather difficult quantity.
We will need much more detailed analysis with larger statistics and/or
finer points of simulations to draw definitive information from kurtosis.

%For $\beta=1.60$, the minimum shows very weak volume dependence and
%tends to increase slightly for larger volume instead of decreasing.
%Thus the kurtosis diagnosis may turn down a possibility of the 1st
%order phase transition or one needs much larger lattice volume to be
%in the scaling region.

\subsection{CLB cumulant}
\begin{figure}
  \includegraphics[scale=1.2]{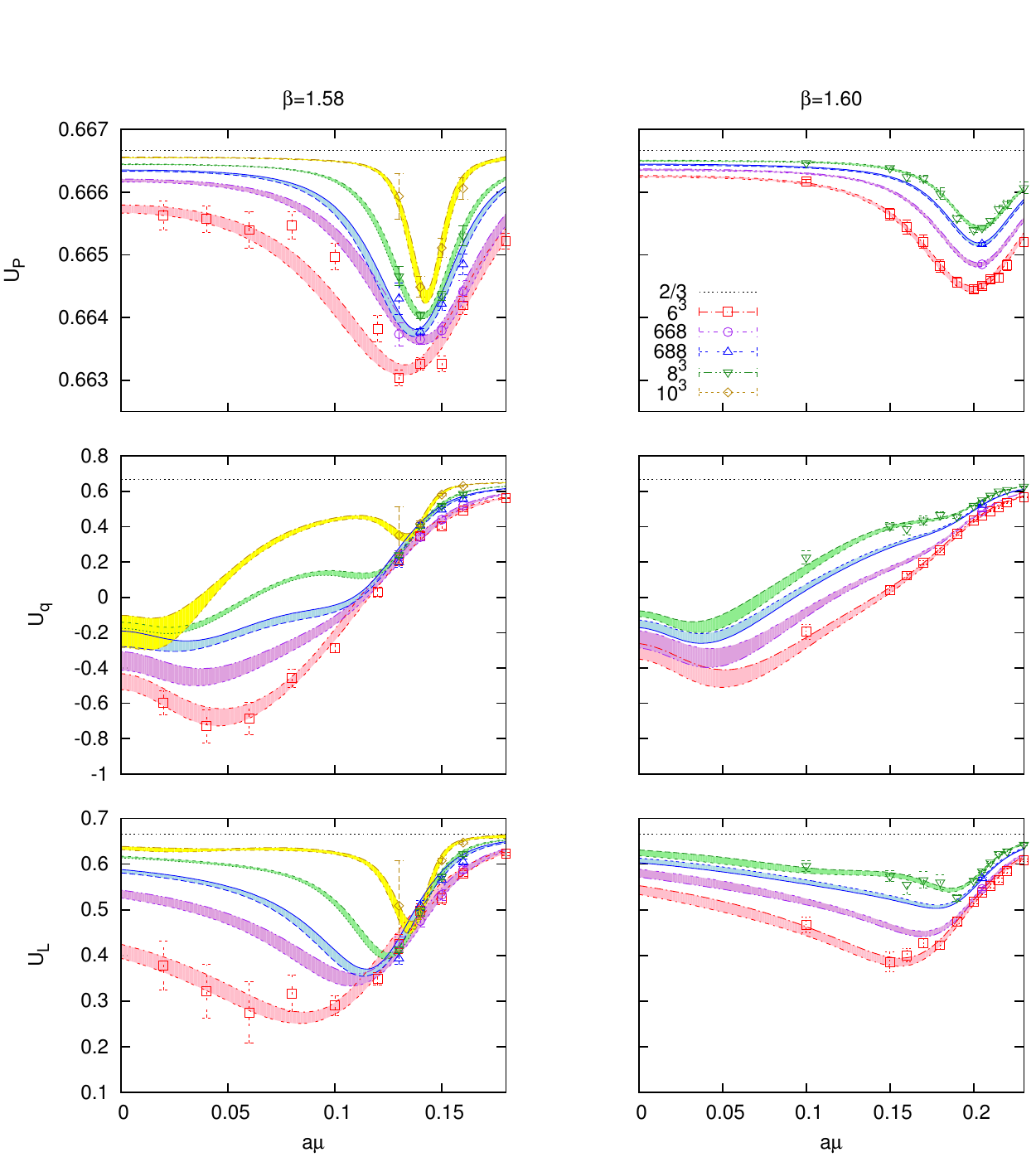}
  \caption{\label{fig:CLBcumulant}CLB cumulant of plaquette $U_P$ (top
    panels), quark number density $U_{\text q}$ (middle panels), and
    the Polyakov loop $U_L$ (bottom panels) as functions of $a\mu$ at
    $\beta=1.58$ (left) and $\beta=1.60$ (right).}
\end{figure}

In Fig.~\ref{fig:CLBcumulant}, we show the CLB cumulant for plaquette
$U_P$, quark number density $U_{\text q}$, and Polyakov loop $U_L$.
Both $U_P$ and $U_L$ show a unique minimum in the region we
investigate.  The volume dependence of the minimum position is rather
large for $U_L$ while it is small for $U_P$.  The results for gauge
action density and fuzzy Polyakov loop show similar trends to that of
plaquette and Polyakov loop, respectively.  In contrast, $U_{\text q}$
exhibits a broad minimum even for relatively large volumes, and there
is an additional minimum generated far away from the transition region
for large volumes.
%{\it ST The definition of $U_{\text q}$ is something wrong? Or, the
%  reweighting is not working in the region $0<a\mu<0.1$ for larger
%  volume? ST}
Since the CLB cumulant is defined in terms of non-central moments, it
may depend more on the detailed form of observable distributions than
those defined in terms of central moments and their ratios.  In any
case we need more understanding on the behavior of $U_{\text q}$, and
we choose not to perform the volume scaling analysis for $U_{\text q}$
in the following.

\begin{figure}
  \subfloat[]{\includegraphics{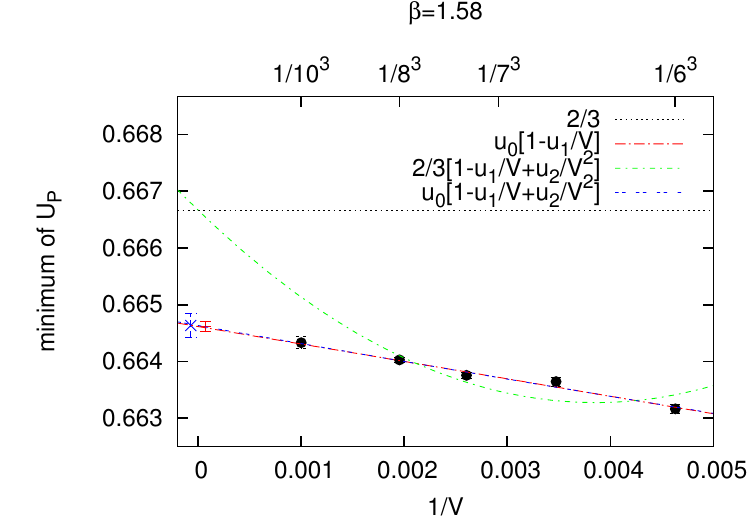}} \qquad
  \subfloat[]{\includegraphics{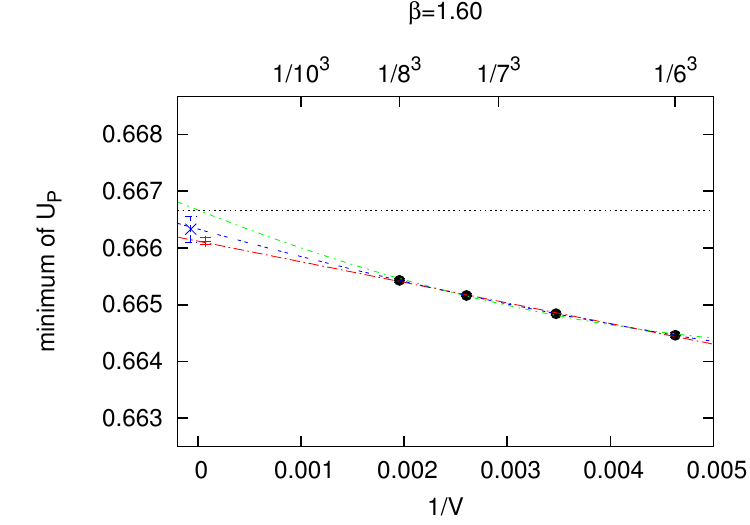}} \\
  \subfloat[]{\includegraphics[scale=0.95]{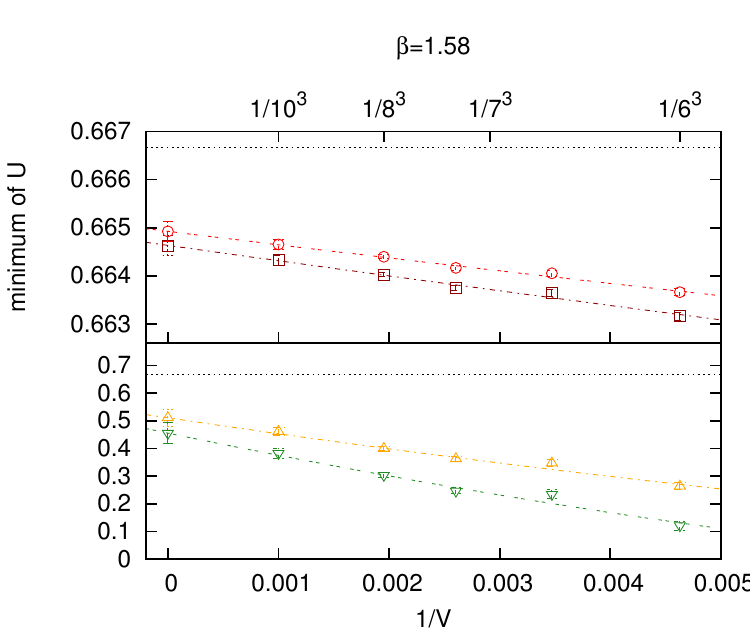}} \qquad
  \subfloat[]{\includegraphics[scale=0.95]{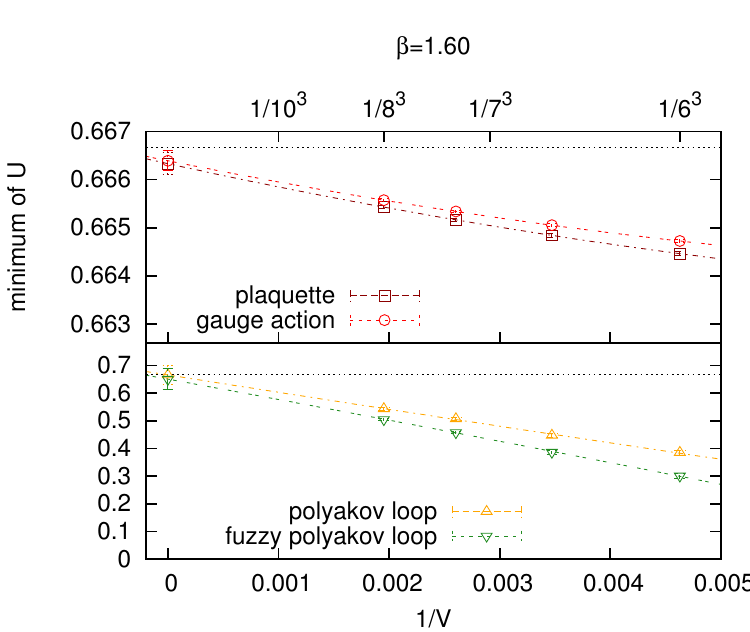}}
  \caption{\label{fig:CLBcumulant2}Upper panels show volume scaling of
    the minimum of $U_P$ for $\beta=1.58$ (left panel) and $1.60$
    (right panel) together with results of three types of fits.  Lower
    panels show volume scaling plot for plaquette, gauge action
    density, Polyakov loop, and fuzzy Polyakov loop, together with
    results of the fitting form C3.}
\end{figure}

In order to extract the infinite volume limit, we perform fitting with
the form
\begin{equation}
  U^{\min}_P
  =
  u_0(1-u_1/V+u_2/V^2),
  \label{eqn:CLBvolumescaling}
\end{equation}
and consider three cases,
\begin{enumerate}
  \item[C1] assuming $u_2=0$
  \item[C2] assuming $u_0=2/3$
  \item[C3] no constraint
\end{enumerate}
The results for fit parameters are summarized in
Table~\ref{tab:u_b1.58} and \ref{tab:u_b1.60} for $\beta=1.58$ and
$1.60$, respectively.  In Fig.~\ref{fig:CLBcumulant2}, the top panels
shows the volume dependence of the minimum value of the CLB cumulant
for plaquette, together with the curves of the three fits.  The
bottoms panels summarize the minimum values for all observable we
consider and the fit curves from the fitting form C3.

We find the results of fits to be essentially the same in character to
those for the susceptibilities.  At $\beta=1.58$, data are well
described by either the fitting form C1 or C3, with consistent values
of the fit parameters.  In particular, $u_0$ clearly deviates away
from $2/3$. On the other hand, the fitting form C2 with $u_0$ fixed at
$2/3$ has an unacceptably large $\chi^2$.  Thus a crossover is
strongly excluded. At $\beta=1.60$, the fitting form C1 and C2 are
equally reasonable.  It is difficult to distinguish between a first
order phase transition and a crossover from present data alone.

%If the $u_0$ in the fitting form 3 is significantly different from $2/3$
%or the fitting form 1 describes the data well then
%the 1st order phase transition is suggested.
%On the other hand, if the $u_0$ in the fitting form 3 is consistent with $2/3$ or
%the fitting form 2 describes the data well then the crossover is indicated.
%The resulting fit parameters are summarized in Table~\ref{tab:u_b1.58}
%and \ref{tab:u_b1.60} for $\beta=1.58$ and $1.60$ respectively.  The
%fitting results for $\beta=1.58$ are well described by $1/V$ scaling
%and the thermodynamic value is significantly different from $2/3$,
%therefore 1st order phase transition is strongly suggested.  On the
%other hand for $\beta=1.60$, it is hard to distinguish between 1st
%order phase transition and crossover.  One neeeds larger volume to
%draw a clear statement about the strengh of the transition.

\begin{table}
  \caption{\label{tab:u_b1.58}Values of fit parameters of volume
    scaling form for CLB cumulant in Eq.~\eqref{eqn:CLBvolumescaling}
    at $\beta=1.58$. Values without error means that the corresponding
    parameter is fixed.}
  \begin{ruledtabular}
    \begin{tabular}{l|c|l|l|l|l}
      observable          & fitting form & $u_{0}$        & $u_{1}$      & $u_{2}$                & $\chi^2/{\text{dof}}$ \\
      \hline
                          & C1 & $0.664614(82)$ & $0.462(44)$  & $0$                    & $1.02$               \\
      plaquette           & C2 & $2/3$          & $2.640(52)$  & $342(15)$              & $31.7$               \\
                          & C3 & $0.66463(21)$  & $0.48(23)$   & $4(38)$                & $1.53$               \\
      \hline
                          & C1 & $0.664901(74)$ & $0.398(40)$  & $0$                    & $0.742$              \\
      gauge action        & C2 & $2/3$          & $2.258(46)$  & $291(13)$              & $26.3$               \\
                          & C3 & $0.66492(20)$  & $0.42(22)$   & $4(35)$                & $1.11$               \\
      \hline
                          & C1 & $0.499(12)$    & $100.3(6.6)$ & $0$                    & $1.39$               \\
      polyakov loop       & C2 & $2/3$          & $255.6(7.4)$ & $0.296(22 )\times10^5$ & $10$                 \\
                          & C3 & $0.511(30)$    & $116(37)$    & $0.30(71 )\times10^4$  & $2$                  \\
      \hline
                          & C1 & $0.435(16)$    & $153.5(8.3)$ & $0$                    & $2$                  \\
      fuzzy polyakov loop & C2 & $2/3$          & $353(10)$    & $0.403(29 )\times10^5$ & $12.1$               \\
                          & C3 & $0.455(38)$    & $182(47)$    & $0.61(99 )\times10^4$  & $2.83$               \\
    \end{tabular}
  \end{ruledtabular}
\end{table}

\begin{table}
  \caption{\label{tab:u_b1.60}Values of fit parameters of volume scaling form
    for CLB cumulant at $\beta=1.60$.}
  \begin{ruledtabular}
    \begin{tabular}{l|c|l|l|l|l}
      observable          & fitting form & $u_{0}$        & $u_{1}$      & $u_{2}$               & $\chi^2/{\text{dof}}$ \\
      \hline
                          & C1 & $0.666118(60)$ & $0.545(30)$  & $0$                   & $0.453$              \\
      plaquette           & C2 & $2/3$          & $1.083(32)$  & $81.2(8.9)$           & $1.05$               \\
                          & C3 & $0.66633(23)$  & $0.76(23)$   & $33(35)$              & $0.00105$            \\
      \hline
                          & C1 & $0.666175(54)$ & $0.478(27)$  & $0$                   & $0.573$              \\
      gauge action        & C2 & $2/3$          & $0.961(29)$  & $73.0(8)$             & $0.869$              \\
                          & C3 & $0.66639(21)$  & $0.69(20)$   & $33(31)$              & $1.02\times10^{-5}$  \\
      \hline
                          & C1 & $0.6621(91)$   & $91.3(3.7)$  & $0$                   & $0.551$              \\
      polyakov loop       & C2 & $2/3$          & $95.4(4.8)$  & $0.08(15)\times10^4$  & $0.545$              \\
                          & C3 & $0.666(34)$    & $95(30)$     & $0.06(54)\times10^4$  & $1.09$               \\
      \hline
                          & C1 & $0.656(11)$    & $117.2(4)$   & $0$                   & $0.164$              \\
      fuzzy polyakov loop & C2 & $2/3$          & $125.5(5.6)$ & $0.15(17)\times10^4$  & $0.234$              \\
                          & C3 & $0.651(39)$    & $112(34)$    & $-0.10(65)\times10^4$ & $0.306$              \\
    \end{tabular}
  \end{ruledtabular}
\end{table}

\subsection{\label{sec:transitionpoint}Transition point}

\begin{figure}
  \subfloat[]{\includegraphics{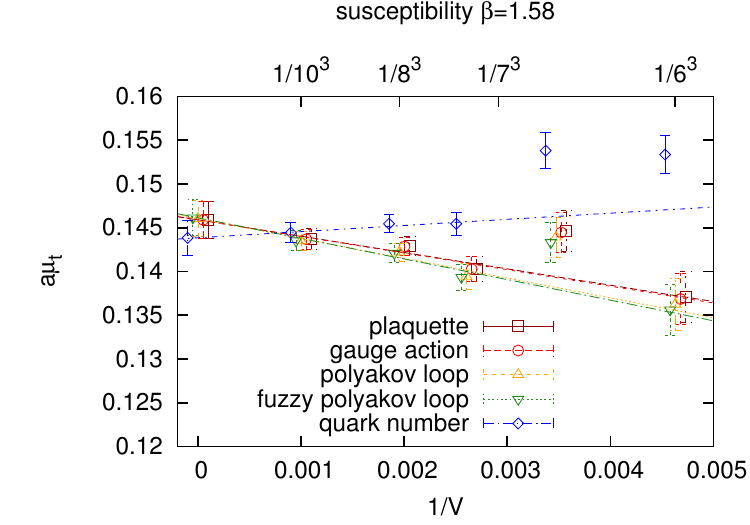}} \qquad
  \subfloat[]{\includegraphics{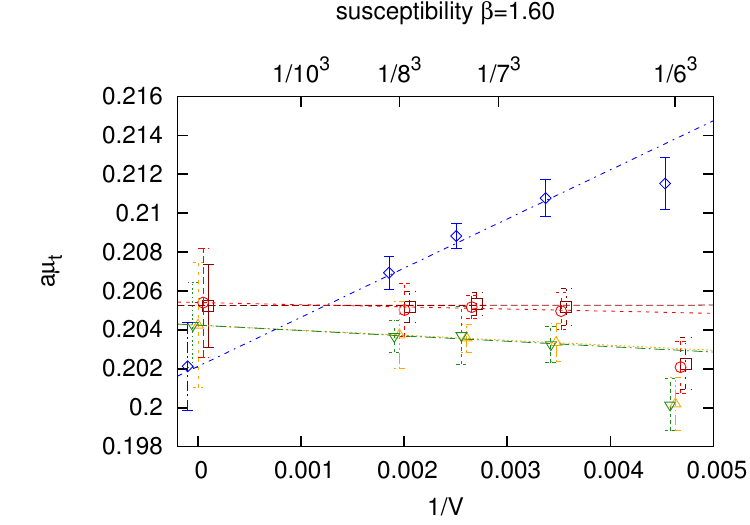}} \\
  \caption{\label{fig:transitionpointfinitemu}Volume dependence of the
    transition point $a\mu_{\text t}$ determined from susceptibility
    peak of several observables for (a) $\beta=1.58$ and (b) $1.60$
    together with the fitted line of
    Eq.~\eqref{eqn:transitionpointvolumescaling}.  }
\end{figure}

The transition point can be determined by the peak of the
susceptibility or the zero of the skewness for each volume.  The
transition point in the infinite volume may then be obtained by a
volume extrapolation with a fitting form
\begin{equation}
  a\mu_{\text t}(V)=a\mu_{\text t}(V=\infty)+A/V,
  \label{eqn:transitionpointvolumescaling}
\end{equation}
where $a\mu_{\text t}(V=\infty)$ and $A$ are fitting parameters.
The volume dependence of the transition point determined from the
susceptibility for five observables, and the volume extrapolation
using Eq.~\eqref{eqn:transitionpointvolumescaling}, are shown in
Fig.~\ref{fig:transitionpointfinitemu}.  The largest three volumes are
used for the fits, namely $V=688,8^3,10^3$ for $\beta=1.58$ and
$V=668,688,8^3$ for $\beta=1.60$.  The transition points determined
from several observables are different from each other at finite
volumes.  However, after taking the infinite volume limit, they
coincide with each other within the estimated errors.  The transition
point determined by the zero of skewness gives the same value within
error at each finite volume, and the final value and the size of error
are similar to those calculated from susceptibilities.  For future
reference we quote the transition point determined from the
susceptibility of plaquette,
\begin{equation}
a\mu_{\text t}(V=\infty)
=
\left\{
\begin{array}{cc}
0.1459(20)& \mbox{ for } \beta=1.58,\\
0.2053(21)& \mbox{ for } \beta=1.60.\\
\end{array}
\right.
\end{equation}

\subsection{\label{sec:pressure}Pressure}

For the grand canonical ensemble approach,
the pressure is given by the corresponding partition function,
\begin{equation}
  p_{\text{QCD}}(\mu)=\frac{T}{V}\ln\mathcal{Z}_{\text{QCD}}(\mu).
\end{equation}
The ratio of two partition functions is thus directly related to their
difference in pressure.
%%(ukawa: I am not sure if this is an explanation of the dip of <cos\theta> since the reason why full pressure and phase quenched pressure differs at the dip is not really explained.  It seems to me that one is just translating one phenomenon in the language of another? ST: Yes indeed. This is just an observation. I thought some reader may be interested in.
%%% YN
%The average phase-reweighting factor, which is the ratio of full QCD
The averaged phase-reweighting factor, which is the ratio of full QCD
%%% YN
partition function and phase-quenched partition function, can be
expressed as the difference in pressure,
\begin{equation}
\langle
\cos(4\theta)
\rangle_{||}
=
\exp\left[\frac{V}{T}\left(p_{\text{QCD}}(\mu)-p_{\text{QCD}_{||}}(\mu)\right)\right]
=
\exp\left[\frac{V}{T}\Delta p(\mu)\right]\le 1.
\label{eqn:pressurediffcos}
\end{equation}

\begin{figure}
  \subfloat[]{\includegraphics{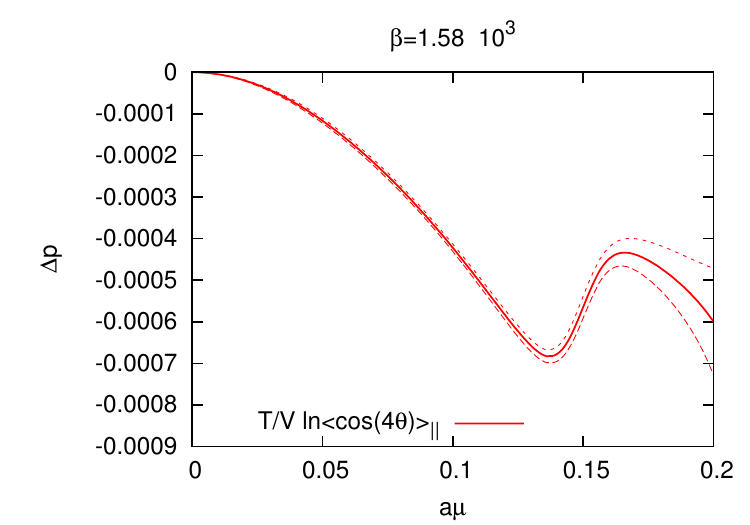}} \qquad
  \subfloat[]{\includegraphics{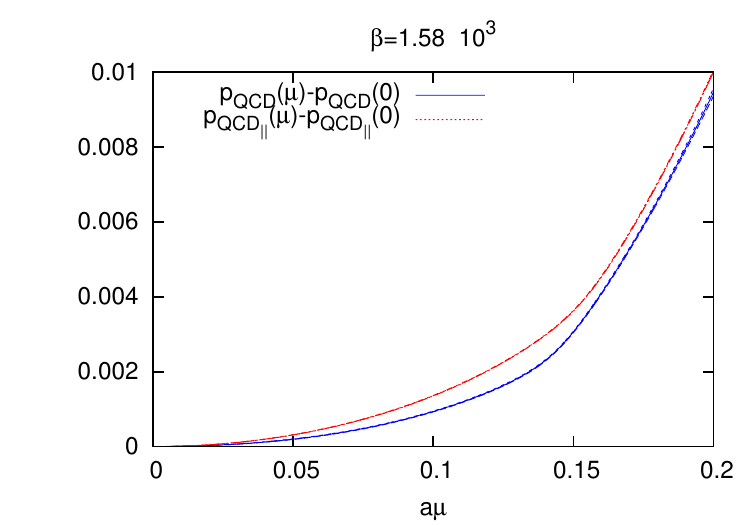}} \\
  \caption{\label{fig:pressurediff}The left figure is the difference
    of the pressure between full QCD and phase quenched QCD with $1\sigma$ error band.
    % (JIN: I suggest removing the result from integrating. ST: removed)
    The right figure is the subtracted pressure for QCD and phase
    quenched QCD as a function of $a\mu$ at $\beta=1.58$ on $10^3$
    lattice.  The inequality $p_{\text{QCD}}(\mu)\le
    p_{\text{QCD}_{||}}(\mu)$ in the shown range of $\mu$ is seen.
    The pressure here is in lattice unit.}
\end{figure}

Conversely, the pressure difference between full QCD and
phase-quenched is given by $T/V \ln \langle\cos(4\theta)\rangle_{||}$,
and is shown in Fig.~\ref{fig:pressurediff}(a).
% JIN: I think showing one method is enough. ST: removed.
This can be compared with Fig.~\ref{fig:phasereweighting}, where the
dip in the phase-reweighting factor manifests itself as the dip in the
pressure difference.  In order to better understand the local minimum,
we compare the pressure from full QCD and phase-quenched directly by
plotting them together in Fig.~\ref{fig:pressurediff}(b).  In this
figure, we show the value of each pressure at chemical potential,
$\mu$, relative to the value at $\mu=0$.  These are computed by
numerically integrating the quark number density
(Eq.~\eqref{eqn:quarknumber}),
\begin{align}
  p(\mu)-p(0)
  &= \frac{T}{V}
  \int_0^\mu d\mu^\prime
  \frac{\partial\ln\mathcal{Z}(\mu^\prime)}{\partial \mu^\prime}, \\
  &= \int_0^\mu d\mu^\prime n_{\text q}(\mu^\prime). \label{eqn:pQCD1}
\end{align}
We can see, in Fig.~\ref{fig:pressurediff}, that there is a change of
slope in full QCD appears at a relative smaller chemical potential
than the change of slope in phase-quenched QCD does.  This produces
the dip.

\begin{figure}
  \subfloat[]{\includegraphics{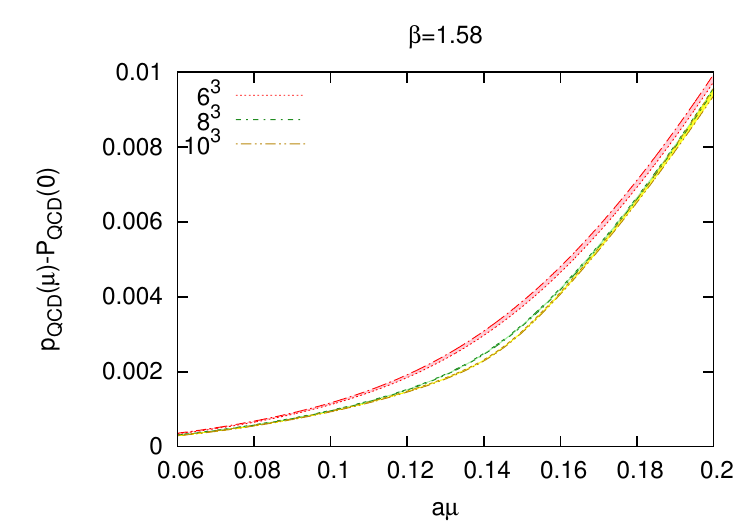}} \qquad
  \subfloat[]{\includegraphics{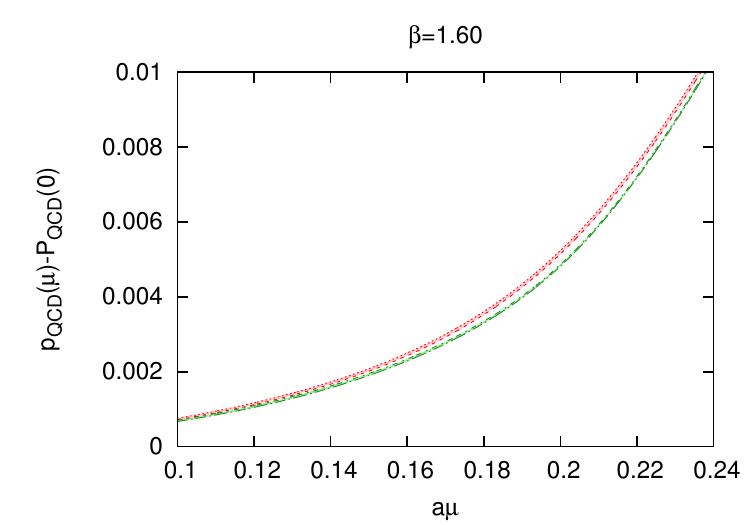}} \\
  \caption{\label{fig:pressure}The left (right) figure is the subtracted pressure as a function of $a\mu$ at
    $\beta=1.58$ ($1.60$). The band shown here is 1 $\sigma$-band.
    }
\end{figure}

The slope in figures of pressure versus chemical potential is quark
number density as given in Eq.~\eqref{eqn:pQCD1}.  The rapid increase
of slope here is the same as a rapid increase of quark number density,
which is an expected behavior for a phase transition.
Fig.~\ref{fig:pressure} shows results of relative pressure in full QCD
from our simulations.  Compared to our moment analysis, at
$\beta=1.58$, where the first order phase transition is suggested, the
slope around the transition point ($a\mu\approx0.146$) changes more rapidly with larger
volumes and it is likely to develop a discontinuity in the first
derivative of pressure in the infinite volume limit, which is a
classical signal of a first order phase transition.  On the other
hand, at $\beta=1.60$ with the volumes we have simulated, the change
is less sharp, which is consistent with results from other moments, namely a crossover.

Finally, after understanding the meaning of the first derivative of
pressure, the dip in Fig.~\ref{fig:pressurediff}(a) can be explained
in the following way.  It appears when the first derivative of
pressure in full QCD changes more rapidly than that in phase-quenched.
When the phase-quenched system is away from a transition while the
full QCD system undergoes a transition, such dip becomes sharper.  The
dip becomes a downward wedge---a discontinuity in slope---in the
thermodynamic limit, when a first order transition occurs.

\section{\label{sec:phasediagram}Global picture of phase diargram}

\begin{figure}
  \subfloat[]{\includegraphics{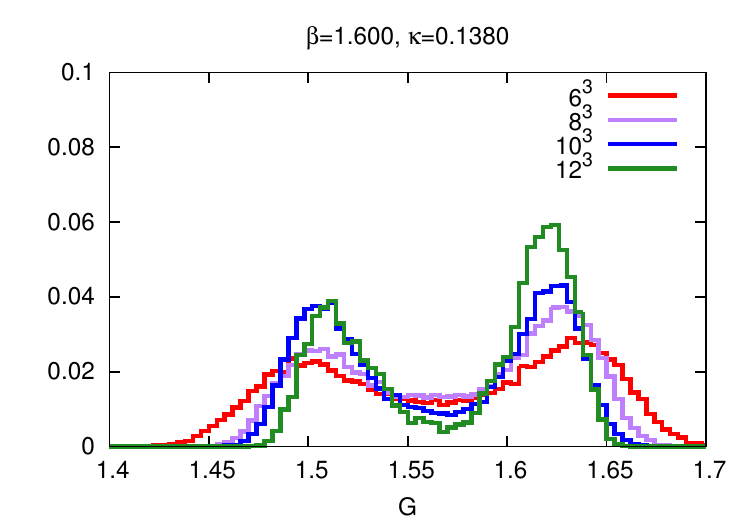}} \qquad
  \subfloat[]{\includegraphics{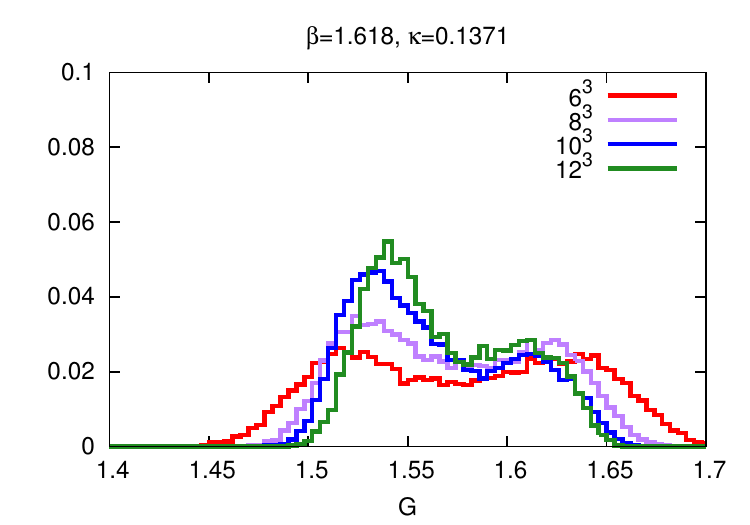}} \\
  \subfloat[]{\includegraphics{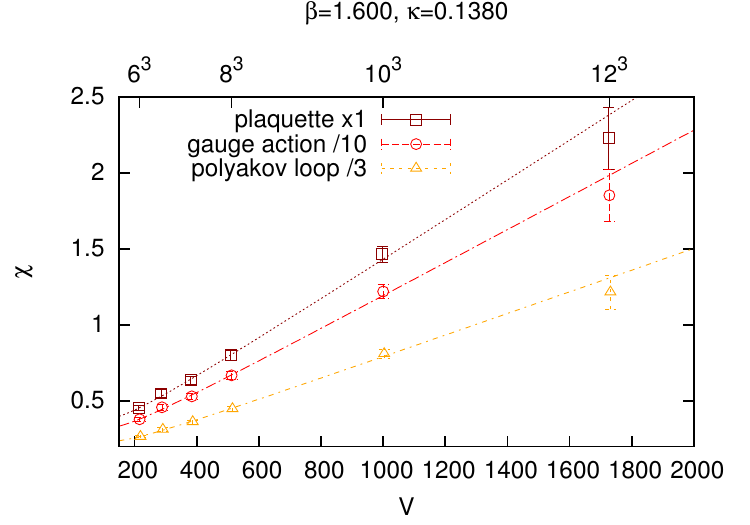}} \qquad
  \subfloat[]{\includegraphics{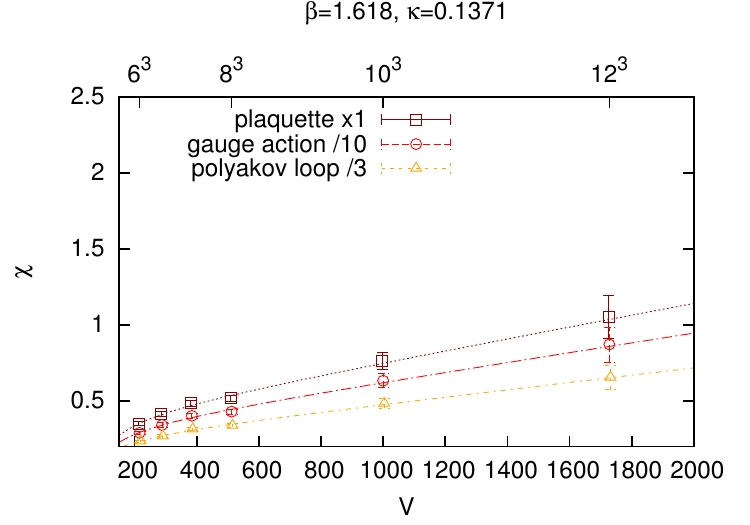}} \\
  \subfloat[]{\includegraphics{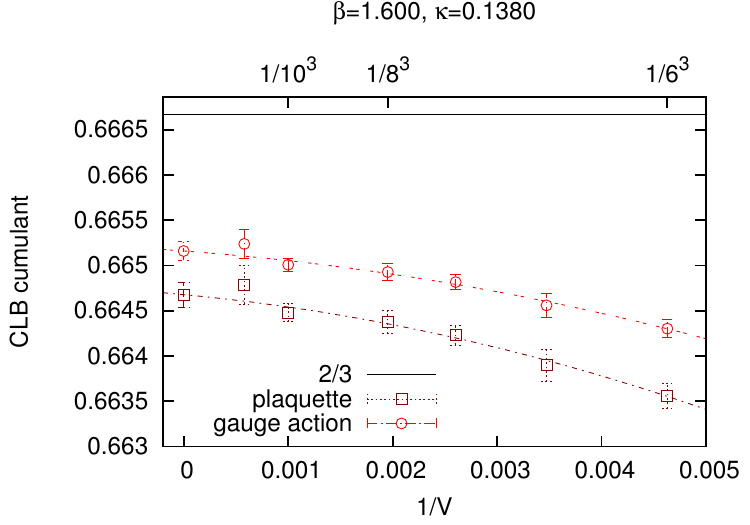}} \qquad
  \subfloat[]{\includegraphics{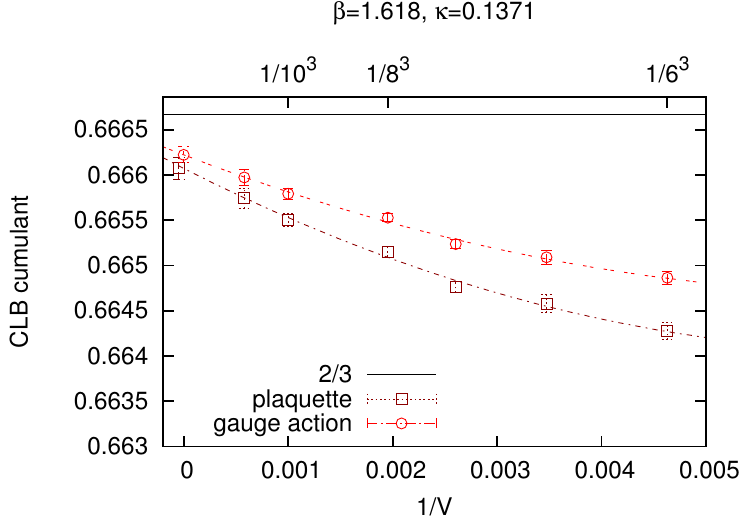}} \\
  \caption{\label{fig:zerodensity}The top figure is the normalized
    histogram of the gauge action density.  The left figures are for
    $\beta=1.600$ and $\kappa=0.1380$ while the right figures are for
    $\beta=1.618$ and $\kappa=0.1371$.  The middle and the bottom
    figures show the volume scaling of susceptibility and the CLB
    cumulant respectively for the plaquette, the gauge action density
    and the Polyakov loop.  In the bottom figures, the CLB cumulant of
    the Polyakov loop is not shown because its minimum is quite far
    from the simulation point.  The curves show the fitting forms S3
    and C3 defined in the Sec.~\ref{sec:finitedensitymoment}.  }
\end{figure}

We may ask what present results can tell us about the phase diagram
depicted in Fig.~\ref{fig:phasediagram}.  To answer this question we
made additional simulations at $a\mu=0$ with $(\beta, \kappa)=(1.600,
0.1380)$ and $(1.618, 0.1371)$.  The volume scaling of the histogram
for the gauge action density and the susceptibility and the CLB
cumulant shown in Fig.~\ref{fig:zerodensity} indicate that the former
point has a clear first order phase transition, while the latter point
has a much weaker transition, possibly consistent with a crossover.
Linearly connecting the two points yields $\kappa_{\text t}\approx
0.2180-0.0500\beta$ as an estimate of the line of transition.  Since
we wish to draw the phase diagram for a fixed quark mass in physical
units, we calculate $m_\pi/m_\rho$ from
Table~\ref{tab:hadronspectrum}, and find $m_\pi/m_\rho\approx
0.555\beta-0.064$ along the line of transition.  Given
$m_\pi/m_\rho=0.822(3)$ also from Table~\ref{tab:hadronspectrum}, we
estimate that the first order transition at $(\beta, \kappa,
a\mu)=(1.58, 0.1380, 0.1459(20))$ is connected to the point $(1.596,
0.1382, 0)$ where we expect a first order transition from the zero
density runs discussed above.  We come to the conclusion that for
$m_\pi/m_\rho=0.822(3)$ the phase diagram looks like
Fig.~\ref{fig:phasediagram}(a).

A similar estimate starting from  $(1.60, 0.1371, 0.2053(21))$ where
$m_\pi/m_\rho=0.839(2)$ indicates that this point is connected to
$(1.627, 0.1367, 0)$ where the transition is either a weak first order or a
crossover.  There is a possibility that the phase diagram looks like
Fig.~\ref{fig:phasediagram}(b).

\section{\label{sec:concludingremarks}Concluding remarks}

Taken together, the results of our finite size scaling analyses show
that there is a first order phase transition at $\beta=1.58$,
$\kappa=0.1380$ and $a\mu=0.1459(20)$.  On the other hand, for the
Kentucky group's parameter set $\beta=1.60$, $\kappa=0.1371$, our
range of lattice sizes from $6^3$ to $8^3$ is not large enough to draw
a clear conclusion about the nature of the transition, although we
have confirmed that the transition point $a\mu_{\text
  t}\approx0.2053(21)$ is very close to that determined by their
canonical approach.

Together with additional zero density simulations, we come to the
conclusion that for $m_\pi/m_\rho=0.822(3)$ the phase diagram looks
like Fig.~\ref{fig:phasediagram}(a).  On the other hand,
$m_\pi/m_\rho=0.839(2)$ indicates that the tansition is either a weak
first order or a crossover and there is a possibility that the phase
diagram looks like Fig.~\ref{fig:phasediagram}(b).

%We will soon report on the partition function zero analysis
%by using multi-ensemble reweighting technique.
%We are currently working on the phase structure of
%$N_{\text f}=3$ QCD with both zero and non-zero desity.

\section*{Acknowledgments}

The authors gratefully acknowledges the useful conversation with Mike
Endress, Sinya Aoki, Kazuyuki Kanaya and Shinji Ejiri.
% We thank Ken-Ichi Ishikawa for providing us his code partially used
% in this work.
This work is supported in part by the Grants-in-Aid for Scientific
Research from the Ministry of Education, Culture, Sports, Science and
Technology (Nos.  23105707, %takeda shingakujutsu
23740177, %takeda wakate B
22244018, %kura kiban A
20105002). %kura shingakujyutsu
% A part of this research has been funded by MEXT HPCI STRATEGIC
% PROGRAM.
The numerical calculations have been done on T2K-Tsukuba and HA-PACS
cluster system at University of Tsukuba.  We thank the Galileo Galilei
Institute for Theoretical Physics for the hospitality and INFN for
partial support offered to S.T. during the workshop ``New Frontiers in
Lattice Gauge Theories'', while this work was completed.

\appendix

\section{\label{sec:doublepeakdistribution}Volume scaling of higher moments
in a double Gaussian model}

%\begin{figure}
%  \includegraphics[scale=1.5]{figure/doublepeak}
%  \caption{\label{fig:doublepeak}Double Gaussian distribution.}
%\end{figure}

In this appendix, we summarize a phenomenological distribution
argument originally due to Ref.~\cite{Challa:1986sk}.  Close to a first order transition point, the distribution of an
observable $X$ can be approximately described by a double Gaussian
form given by
\begin{equation}
  P(X)
  =
  a_+ \sqrt{\frac{V}{2\pi c_+}} e^{-\frac{(X-x_+)^2}{2c_+/V}}
    +
  a_- \sqrt{\frac{V}{2\pi c_-}} e^{-\frac{(X-x_-)^2}{2c_-/V}}
  .
  \label{eqn:doublepeakdistribution}
\end{equation}
%%ukawa: figure is probably not needed.
%as shown in Fig.~\ref{fig:doublepeak}. ST: I remove it.
This distribution is normalized
\begin{equation}
  \int^\infty_{-\infty}
  P(X)dX=1,
\end{equation}
provided $a_+ + a_- = 1$.  Any observable $f(X)$ of $X$ can be
calculated as
\begin{equation}
  \langle f(X)\rangle=\int_{-\infty}^{\infty} \dd X f(X) P(X).
\end{equation}

Let $t$ be the parameter controlling the phase transition, {\it e.g.},
temperature, and let $a_+=a_-=1/2$ or $t=0$ be the transition point at
infinite volume.  The infinite volume free energy density has two
branches which cross at $t=0$, and switches the minimum.  Normalizing
the scale of $t$, one can write
\begin{equation}
\label{eq:def-a}
a_\pm = \frac{e^{\pm Vt}}{e^{Vt}+e^{-Vt}}.
\end{equation}

Simple but tedius calculation leads to the following expressions for
the susceptibility, skewness, kurtosis, and the CLB cumulant:
\begin{align}
  \chi_X &= V \langle (X-\langle X\rangle)^2 \rangle
  = Va_+a_-(x_+-x_-)^2 + (a_+c_+ + a_-c_-),
  \\
  S_X & = \frac{ \langle (X-\langle X\rangle)^3 \rangle }{ \langle
    (X-\langle X\rangle)^2 \rangle^{3/2} }
  = - \frac{a_+-a_-}{\sqrt{a_+a_-}}+O(V^{-1}),
  \\
  K_X &= \frac{ \langle (X-\langle X\rangle)^4 \rangle }{ \langle
    (X-\langle X\rangle)^2 \rangle^{2} } -3
        = -2 +\frac{1-4a_+a_-}{a_+a_-}+O(V^{-1}),
  % \label{eqn:kurtosisscaling}
  \\
  U_X &= 1-\frac{1}{3} \frac{ \langle X^4 \rangle }{ \langle X^2
    \rangle^2 }
        = \frac{2}{3} - \frac{a_+a_-(x_+^2-x_-^2)^2}{3 (a_+x_+^2+a_-x_-^2)^2}
        +O(V^{-1}).
\end{align}
Another simple calculation of derivative with respect to $t$ leads to
\begin{align}
  \frac{d\chi_X}{dt} &= - b(a_+-a_-)(x_+-x_-)^2V^2 + b(c_+-c_-)V,
    \\
  \frac{dS_X}{dt} & = -\frac{b}{2(a_+a_-)^{3/2}}V + O(V^0),
  \\
  \frac{dK_X}{dt} &= - b\frac{a_+-a_-}{(a_+a_-)^2}V + O(V^0),
  \\
  \frac{dU_X}{dt} &= \frac{b}{3} \frac{(a_+x_+^2-a_-x_-^2)}{(a_+x_+^2+a_-x_-^2)^3}
   (x_+^2-x_-^2)^2V +O(V^0),
\end{align}
with $b=2/(e^{Vt}+e^{-Vt})$.

From the above equations, we read that the peak of susceptibility,
zero of skewness and minimum of kurtosis take place at the same value
$t=0$ up to corrections of $O(V^{-2})$.  Expanding the skewness and
kurtosis in the leading orders of $V$ around $t=0$ with $Vt\ll1$, we find
\begin{align}
  S_X & = - 2Vt + O(V^0),
  \\
  K_X &= -2 + 4V^2t^2+ O(V^1).
\end{align}
Therefore, in the leading order, the slope of skewness increases
linearly, and the curvature of kurtosis quadratically, with volume.

The CLB cumulant exhibits a subtlety.  The minimum position deviates from
$t=0$ by $O(V^{-1})$:
\begin{equation}
t_{\text{CLB min}}=\frac{1}{2V}\ln\frac{x_-^2}{x_+^2}+O(V^{-2})
\end{equation}
The infinite volume values at this minimum and at $t=0$ differ:
\begin{align}
\left. U_X\right|_{t=t_{\text{CLB min}}} & = \frac{2}{3}-\frac{(x_+^2-x_-^2)^2}{12x_+^2x_-^2} + O(V^{-1}),\\
\left. U_X\right|_{t=0} & = \frac{2}{3}-\frac{(x_+^2-x_-^2)^2}{3(x_+^2+x_-^2)^2}
+ O(V^{-1}).
\end{align}
This may seem paradoxical that $\lim_{V\rightarrow\infty}t_{\text{CLB
    min}} = 0$, while $\lim_{V\rightarrow\infty}
\left. U_X\right|_{t=t_{\text{CLB min}}} \neq
\left. U_X\right|_{t=0}$.  This is because, at the minimum of the CLB
cumulant, $\lim_{V\rightarrow\infty}\frac{a_+}{a_-} =
\frac{x_-^2}{x_+^2}$, which is away from unity where the phase
transition occurs even in the infinite volume limit.

\section{\label{sec:mureweighting_for_quark_number}Remark on $\mu$-reweighting
  for quark number related quantities}
Note that the observables, like plaquette value, gauge action,
Polyakov loop, fuzzy Polyakov loop are independent of $\mu$, while the
quark number density has an explicit $\mu$-dependence.  Therefore we
have to identify a difference in the observable
\begin{equation}
  \mathcal{O}(\mu^\prime) =
  \mathcal{O}(\mu) + \Delta \mathcal{O}(\mu^\prime,\mu).
\end{equation}
Before identifying the difference, first let us remind the quark
number related quantities.  Actually, they can be expressed by using
$Q_n$ in Eq.~\eqref{eqn:Q} as follows.
In order to construct the quark number related observable at
$\mu^\prime$ we have to know $Q_n(\mu^\prime)$.  For that purpose, we
have to know $W_n(\mu^\prime/T)$ as seen from Eq.~\eqref{eqn:Q}.
\begin{equation}
  W_n(\mu^\prime/T)
  =
  W_n(\mu/T)
  +
  \Delta W_n(\mu^\prime/T,\mu/T).
\end{equation}
There are two ways to approximate $\Delta W_n(\mu^\prime/T)$, namely
the winding expansion and the Taylor expansion.  In the following we
show only the latter and it is given by
\begin{align}
  W_n(\mu^\prime/T) &= \sum_{m=0}^\infty
  \frac{(\mu^\prime/T-\mu/T)^m}{m!}  \frac{\partial^m
    W_n(\mu/T)}{\partial (\mu/T)^m}
  \nonumber\\
  &= \sum_{m=0}^\infty \frac{(\mu^\prime/T-\mu/T)^m}{m!}
  W_{n+m}(\mu/T),
\end{align}
where we have used a relation
\begin{equation}
  \frac{\partial^m W_n}
  {\partial (\mu/T)^m}
  =
  W_{n+m}.
  \label{eqn:defWn2}
\end{equation}
We truncate the expansion
up to $m=3$ and their explicit forms for $n=1,2,3,4$ are given by
\begin{align}
  W_1(\mu^\prime/T) &= W_1 + (\mu^\prime/T-\mu/T) W_2 +
  \frac{(\mu^\prime/T-\mu/T)^2}{2} W_3 +
  \frac{(\mu^\prime/T-\mu/T)^3}{3!}  W_4,
  \\
  W_2(\mu^\prime/T) &= W_2 + (\mu^\prime/T-\mu/T) W_3 +
  \frac{(\mu^\prime/T-\mu/T)^2}{2} W_4 +
  \frac{(\mu^\prime/T-\mu/T)^3}{3!}  W_5,
  \\
  W_3(\mu^\prime/T) &= W_3 + (\mu^\prime/T-\mu/T) W_4 +
  \frac{(\mu^\prime/T-\mu/T)^2}{2} W_5 +
  \frac{(\mu^\prime/T-\mu/T)^3}{3!}  W_6,
  \\
  W_4(\mu^\prime/T) &= W_4 + (\mu^\prime/T-\mu/T) W_5 +
  \frac{(\mu^\prime/T-\mu/T)^2}{2} W_6 +
  \frac{(\mu^\prime/T-\mu/T)^3}{3!}  W_7.
\end{align}
We approximate $W_5 = W_7 = \tr [B]$ and $W_6 = \tr [C]$.  The error
of this approximation is suppressed by $(\Delta\mu/T)^n/n!$, and is
relatively unnoticeable compared to the statistical error.
%  (This %approximation is empirically not so bad.)

In this way, we obtain the difference
$\Delta W_n(\mu^\prime/T,\mu/T)$ and then from this
one can construct the difference of any quark number related observable.

\bibliography{ref}

\end{document}